\documentclass[iop]{emulateapj} 

\usepackage {epsf}
\usepackage{epstopdf}
\usepackage {lscape, graphicx}
\usepackage{enumitem}
\gdef\1054{MS\,1054--03}

\def\farcs{\hbox{$.\!\!^{\prime\prime}$}}
\def\simgeq{{\raise.0ex\hbox{$\mathchar"013E$}\mkern-14mu\lower1.2ex\hbox{$\mathchar"0218$}}} 
\def\sersic{S\'{e}rsic}
\def\H160{$H_{160}$}

\def\chis{$\chi^{2}$}

\begin {document}

\title {Bulge Growth and Quenching since z = 2.5 in CANDELS/3D-HST}

 \author{Philipp Lang\altaffilmark{1}, Stijn Wuyts\altaffilmark{1}, Rachel S. Somerville\altaffilmark{2}, Natascha M. F{\"o}rster Schreiber\altaffilmark{1}, Reinhard Genzel\altaffilmark{1}, Eric F. Bell\altaffilmark{3}, Gabe Brammer\altaffilmark{4},  Avishai Dekel\altaffilmark{5}, Sandra M. Faber\altaffilmark{7}, Henry C. Ferguson\altaffilmark{8}, Norman A. Grogin\altaffilmark{8}, Dale D. Kocevski\altaffilmark{9}, Anton M. Koekemoer\altaffilmark{8}, Dieter Lutz\altaffilmark{1}, Elizabeth J. McGrath\altaffilmark{10}, Ivelina Momcheva\altaffilmark{6}, Erica J. Nelson\altaffilmark{6}, Joel R. Primack\altaffilmark{11}, David J. Rosario\altaffilmark{1}, Rosalind E. Skelton\altaffilmark{12}, Linda J. Tacconi\altaffilmark{1}, Pieter G. van Dokkum\altaffilmark{6}, Katherine E. Whitaker\altaffilmark{13}}

\altaffiltext{1}{Max-Planck-Institut f\"{u}r extraterrestrische Physik, Giessenbachstrasse, D-85748 Garching, Germany}
\altaffiltext{2}{Department of Physics and Astronomy, Rutgers, The State University of New Jersey, 136 Frelinghuysen Road, Piscataway, NJ 08854, USA}
\altaffiltext{3}{Department of Astronomy, University of Michigan, 500 Church Street, Ann Arbor, MI 48109, USA}
\altaffiltext{4}{European Southern Observatory, Alonson de C\'{o}rdova 3107, Casilla 19001, Vitacura, Santiago, Chile}
\altaffiltext{5}{Center for Astrophysics and Planetary Science, Racah Institute of Physics, The Hebrew University, Jerusalem, 91904, Israel}
\altaffiltext{6}{Astronomy Department, Yale University, New Haven, CT 06511, USA}
\altaffiltext{7}{UCO/Lick Observatory, Department of Astronomy and Astrophysics, University of California, Santa Cruz, CA 95064, USA}
\altaffiltext{8}{Space Telescope Science Institute, 3700 San Martin Drive, Baltimore, MD 21218, USA}
\altaffiltext{9}{Department of Physics and Astronomy, University of Kentucky, Lexington, KY 40506, USA}
\altaffiltext{10}{Department of Physics and Astronomy, Colby College, Waterville, ME 0490, USA}
\altaffiltext{11}{Department of Physics, University of California at SantaCruz, Santa Cruz, CA 95064, USA}
\altaffiltext{12}{South African Astronomical Observatory, Observatory Road, 7925 Cape Town, South Africa}
\altaffiltext{13}{Astrophysics Science Division, Goddard Space Flight Center, Greenbelt, MD 20771, USA}

\begin{abstract}
Exploiting the deep high-resolution imaging of all 5 CANDELS fields,
and accurate redshift information provided by 3D-HST, we investigate
the relation between structure and stellar populations for a
mass-selected sample of 6764 galaxies above $10^{10}\ M_{\sun}$,
spanning the redshift range $0.5<z<2.5$.  For the first time, we fit 2-dimensional models comprising a single \sersic\ fit and two-component (i.e., bulge + disk) decompositions not only to the $H$-band light distributions, but also to the stellar mass maps reconstructed
from resolved stellar population modeling.  We confirm that the increased bulge
prominence among quiescent galaxies, as reported previously based on
rest-optical observations, remains in place when considering the
distributions of stellar mass.  Moreover, we observe an increase of the
typical \sersic\ index and bulge-to-total ratio (with median $B/T$ reaching 40-50\%) among star-forming galaxies above $10^{11}\ M_{\sun}$.  
Given that quenching for these most massive systems is likely to be
imminent, our findings suggest that significant bulge growth precedes
a departure from the star-forming main sequence.  We demonstrate that
the bulge mass (and ideally knowledge of the bulge and total mass) is
a more reliable predictor of the star-forming versus quiescent state
of a galaxy than the total stellar mass.  The same trends are predicted by
the state-of-the-art semi-analytic model by Somerville et al.  In the
latter, bulges and black holes grow hand in hand through merging
and/or disk instabilities, and AGN-feedback shuts off star formation. Further observations will be required to pin down star formation quenching mechanisms, but our results imply they must be internal to the galaxies and closely associated with bulge growth.

\end{abstract}


\section {Introduction}
\label{intro.sec}

The mechanisms driving the shutdown of star-forming galaxies (SFGs), 'quenching', remains one of the least understood puzzles in galaxy formation to date.  In the low-redshift universe, galaxies show a bimodal color distribution, accompanied by a bimodality of morphologies \citep[see][]{kauffmann2003,strateva2001}. Spiral galaxies have low bulge-to-disk ratios and are commonly the site of active star formation leading to blue colors.  Passive galaxies are observed to be mostly  spheroid-dominated.  The color bimodality has also been observed at higher redshifts \citep[e.g.,][]{brammer2009,whitaker2011}.  These observations have been interpreted as evolutionary paths, in which one or several quenching processes cause the SFG to become red and passive on a short timescale \citep [see][]{bell2004,faber2007}.  
Studies of the shape of the mass function of passive galaxies over cosmic time using large surveys such as SDSS, NMBS, zCOSMOS, UltraVISTA and zFOURGE showed that the probability for a galaxy being quenched increases with its mass (Peng et al. 2010; Brammer et al. 2011; Ilbert et al. 2013; Muzzin et al. 2013; Woo et al. 2013; Tomczak et al. 2014).  
Further evidence for quenching also comes from studies employing abundance matching techniques, where the cumulative abundance of galaxies is matched to that of haloes using the results of cosmological dark matter simulations alongside with observational constraints on the stellar mass function.  Those infer low baryon fractions within dark matter haloes, hinting at an efficient quenching mechanism associated with significant gas mass loss for both low and high mass galaxies \citep{moster2010,behroozi2010}.  

Several quenching mechanisms have been proposed, which act to either remove the gas from the galaxy or prevent the existing/inflowing gas within the galaxy to form stars.  For the high-mass regime of galaxies, powerful AGN feedback, which may be induced by galaxy mergers \citep[e.g.,][]{hopkins2006} or internal evolutionary processes triggered by (violent) gravitational disk instabilities \citep[e.g.,][]{bournaud2011}, may drive energetic outflows expelling gas out of the galaxy and heating the halo.
In addition, a so-called 'radio-mode' feedback may suppress the cooling of gas onto the galaxy over a longer timespan (Croton et al. 2006).

Moreover, morphological quenching, proposed by \cite{martig2009} may switch off or reduce the efficiency of star formation in galaxies through the existence of a dominant bulge which stabilizes the gas disk against gravitational instabilities (see Saintonge et al. 2012 and Crocker et al. 2012 for observational hints in the local universe, and Genzel et al. 2014 for galaxies at high redshift).  Another process proposed to be responsible for the shut-down of galaxies is halo mass quenching (e.g., Birnboim \& Dekel 2003; Kere{\v s} et al. 2005).  In this scenario, dark matter haloes 
exceeding a critical mass of $M \sim 10^{12}M_{\odot}$ are able to stop the flow of incoming cold gas onto their central galaxies via virial shock heating, which leads to a decrease of star formation and/or suppresses star formation over longer timescales.  However, at $z>z_{crit}$ (with $ 1 < z_{crit} < 3$), gas is predicted to penetrate to the central galaxy through cold streams even in massive haloes (e.g., Dekel et al. 2009).

The above quenching mechanisms could plausibly leave an imprint on the structure of galaxies, as they may lead to high central concentrations due to internal processes and/or major mergers.  This idea triggered a plethora of studies investigating the relation between star formation activity (or absence thereof) and galaxy structure.  
Local galaxy surveys demonstrated that quiescence is strongly linked with structural and morphological parameters such as the \sersic\ index, stellar mass density and the central velocity dispersion \citep{kauffmann2003,kauffmann2006,schiminovich2007,bell2008,fang2013,cheung2012}.   This suggests that a high stellar mass surface density in the center of a galaxy is connected to its quenching.  

Recently, high-resolution imaging facilities onboard the Hubble Space Telescope (HST), including critically the NIR Wide-Field Camera 3 (WFC3), have allowed us to trace the relation between structure and stellar populations further back in cosmic time.  
Early studies on the basis of high-resolution HST-imaging found a correlation between color and bulge fraction for large samples of galaxies at $z\sim 0.7 - 1$ (Bell et al. 2004; Weiner et al. 2005; Koo et al. 2005).  Particularly the Cosmic Assembly Near-infrared Deep Extragalactic Legacy Survey \citep[CANDELS, ][]{grogin2011,koekemoer2011} has played a pivotal role due to its unique combination of multi-wavelength high-resolution imaging, large sample size, and depth.  A common theme of several of the early CANDELS studies is that a correlation between galaxy structure and stellar populations (a `Hubble sequence') has already been in place since at least $z \sim 2.5$ (Wuyts et al. 2011; Bell et al. 2012; Wang et al. 2012; Lee et al. 2013; see also Cheung et al. 2012).  
Additionally, Bruce et al. (2012) show that the majority of star-forming galaxies have disk-dominated light profiles and passive galaxies have bulge-dominated profiles directly from decomposed bulge+disk light-profile fitting of $M > 10^{11}M_{\odot}$ galaxies at $1 < z < 3$.  However, it should be noted that they also find a significant fraction ($\sim 30 \%$) of passive disk-dominated galaxies and star-forming bulge-dominated systems in this high-mass regime.  
These results have set on a firmer footing the pioneering studies by \cite{franx2008} and \cite{kriek2009}, that used ground-based data and significantly smaller samples, respectively.  At all epochs since at least $z \sim 2.5$, SFGs follow a relatively tight relation of $0.2-0.3$ dex scatter between their ongoing star formation rate (SFR) and assembled stellar mass, with a zero point that increases with lookback time.  This relation has been dubbed the 'main sequence' (MS) of star-forming galaxies (Noeske et al. 2007; Elbaz et al. 2007; Daddi et al. 2007; Rodighiero et al. 2011).  Typical SFGs that lie on the MS are best approximated by exponential disk profiles, while the quiescent galaxy population below the MS is better described with higher \sersic\ indices \citep{sw2011b,bell2012}.  In common to all of the above studies, however, is that they were carried out on monochromatic (albeit mostly rest-optical) imaging, furthermore often exploiting only a subset of the entire CANDELS data set.  If a galaxy 
component (e.g., disk) 
forms 
significantly more 
stars than another component (e.g., bulge), their relative weight in light can differ significantly from their contributions in physically more relevant  units of stellar mass. 

Deep panchromatic high-resolution imaging datasets enable us to go beyond measurements of light profiles, by reconstructing the distribution of stellar mass (see, e.g., Zibetti et al. 2009 for a detailed application to nearby galaxies).  While pioneering work on pre-existing high-z datasets using the NICMOS camera onboard HST was limited to small sample sizes \citep[e.g.,][]{elmegreen2009,fs2011}, \cite{sw2012} made use of resolved 7-band stellar-population synthesis modeling on a pixel-by-pixel basis of over 600 massive ($M > 10^{10}M_{\odot}$) SFGs at $0.5 < z < 2.5$ and found that galaxies are overall smoother and more centrally concentrated in mass than they appear in light.  Caused by a combination of internal extinction and star formation history variations \citep[see also][] {guo2012,lanyon-foster2012,szomoru2013,sw2013}, the presence of mass-to-light ratio variations within galaxies emphasizes the importance of making measurements on mass maps when characterizing the stellar structure of high-
redshift galaxies. Guo et al. (2011) furthermore demonstrated that also quiescent galaxies at high redshift feature color (and hence mass-to-light ratio) gradients.

In order to further shed light on the connection between galaxy structure and quenching, we reconstruct stellar mass maps of a large mass-selected sample ($> 10^{10}\ M_{\sun}$) of galaxies at $0.5 < z < 2.5$ and subject those to a detailed structural analysis.  For this purpose, we exploit the available multi-wavelength data in all 5 CANDELS/3D-HST fields with accurate redshifts from the 3D-HST survey \citep{brammer2012,vd2011}.  We don't restrict the structural measurements to one-component (\sersic) fits, but rather also explore two-component (bulge + disk) decompositions.  The latter allows us to carry out a more direct comparison to semi-analytic models (SAMs; specifically those by Somerville et al. 2008, 2012, further developed by Porter et al. 2014), whose prescriptions are formulated in units of disk and bulge components.  

The paper is structured as follows.  In Section\ \ref{obs_sample.sec}, we give an overview of the observations and sample.  The construction of stellar mass maps through resolved stellar population modeling, and the methodology to derive structural parameters, is described in Section\ \ref{method.sec}.  We present our observational results in Section\ \ref{structure.sec}, followed by a comparison with SAMs in Section\ \ref{SAM_results.sec}.  Finally, we summarize and discuss the implications of our findings in Section\ \ref{discussion.sec}.

Throughout this paper, we quote magnitudes in the AB system, assume a Chabrier (2003) initial mass function (IMF), and adopt the cosmological parameters $(\Omega _M, \Omega _{\Lambda}, h) = (0.3, 0.7, 0.7)$.

\section{Data and Sample selection}
\label{obs_sample.sec}

The core dataset used for this work is the deep space-based HST imaging from the CANDELS multi-cycle treasury program \citep{grogin2011, koekemoer2011}, complemented with redshift information from the 3D-HST grism survey (van Dokkum et al. 2011; Brammer et al. 2012)\footnote{The AGHAST G141 grism observations within GOODS-North from the GO-11600 program (PI: B. Weiner) are included in the analysis under `3D-HST' (Brammer et al. 2012).}.  

In order to determine galaxy-integrated masses and SFRs, we make use of additional multi-wavelength data in the CANDELS/3D-HST fields including space-based photometry from {\it Spitzer}/IRAC, {\it Spitzer}/MIPS and {\it Herschel}/PACS and an array of ground-based facilities (see Skelton et al. 2014 for a detailed description of the $H$-selected multi-wavelength catalogs produced by the 3D-HST team).  

The galaxy-integrated properties are derived following identical procedures as \cite{sw2011b}.  These include stellar masses, based on $U$- to - $8\mu m$ broad-band SED modeling using population synthesis models from \cite{bc03} and SFRs derived from a `ladder of SFR indicators'.  The latter method uses detected emission in either UV + PACS for PACS-detected galaxies (Lutz et al. 2011; Magnelli et al. 2013) or UV + MIPS 24$\mu m$ for MIPS-detected galaxies to compute the sum of the obscured and unobscured SFR. For galaxies lacking an IR detection, the SFR is adopted from the best-fit SED model.  
We verified that consistent results are obtained when splitting star-forming and quiescent galaxies based on the above measures of star formation activity or when adopting the rest-UVJ diagram (Wuyts et al. 2007; Williams et al. 2009)

\subsection{HST imaging}
\label{fields_data.sec}

The HST CANDELS observations used for this study comprise high-resolution imaging in five distinct fields: GOODS-South, GOODS-North, COSMOS, UDS, and EGS, covering a total area of 625 arcmin$^2$.  Typical limiting depths in \H160 are 27.0 mag for CANDELS/wide and 27.7 mag for CANDELS/deep (the central halves of the GOODS fields), respectively.  Additional data used for this work include pre-existing ACS imaging in the GOODS, EGS, and COSMOS fields (Giavalisco et al. 2004; Davis et al. 2007; Koekemoer et al. 2007), plus WFC3 imaging in ERS \citep{windhorst2011}.
The available passbands are $B_{435}$,$V_{606}$,$i_{775}$,$z_{850}$,$J_{125}$,$H_{160}$ for the GOODS fields and $V_{606}$,$I_{814}$,$J_{125}$,$H_{160}$ for the remaining fields. 
Additionally, $Y_{098}$ imaging was available for ERS as part of GOODS-South and $Y_{105}$ for the regions with CANDELS/Deep coverage.

All imaging used in our analysis was drizzled to a $0\farcs06$ pixel scale.  For details on the observations and data reduction, we refer the reader to \cite{koekemoer2011} and \cite{grogin2011}.

\subsection{Sample definition}
\label{fields_data.sec}

The main focus of this paper is on the structural shape of stellar mass distributions in a mass-selected sample of galaxies in the redshift range $ 0.5 < z < 2.5$. For this purpose, we only apply a mass cut  $M > 10^{10}M_{\odot}$ to select our galaxies, well above the mass completeness limit of the $H_{160}$-selected catalogs for all five fields.  Our sample consequently spans a wide range of SFRs, from normal SFGs on the main sequence to quiescent galaxies (QGs) below, that already formed the bulk of their stars.  For selection, we use the SED modeled galaxy-integrated mass estimates, but note that they match well the masses obtained by summing the resolved stellar mass distributions \citep[see also][]{sw2012}.

In order to determine redshifts for the galaxies in our sample, we use ground-based spectroscopic redshift information whenever available. Otherwise, redshifts are fitted to the combination of 3D-HST grism data and broad-band photometry.   

The total sample comprises 6764 galaxies, of which 3839 and 2925 lie within the redshift range $ 0.5 < z < 1.5$ and  $ 1.5 < z < 2.5$, respectively.  In the following, these two redshift ranges are referred to as $z \sim 1$ and $z \sim 2$, respectively.

\section{Methodology}
\label{method.sec}

\subsection{Resolved SED Modeling}
\label{fields_data.sec}

A detailed description of the resolved SED modeling can be found in \cite{sw2012}.  Here, we review only the key steps involved and additional processing steps.

First, all images in the available wavelength bands are brought to the WFC3 \H160 resolution ($0\farcs 18$) by using the IRAF PSFMATCH algorithm.  Next, a Voronoi-binning scheme from \cite{cappellari2003} is applied in order to ensure a minimum S/N level of 10  for each bin in the corresponding $H$-band image. The binned multi-wavelength images are then fit with stellar population synthesis models from \cite{bc03}, assuming a Chabrier (2003) IMF, a uniform solar metallicity, and a Calzetti et al. (2000) reddening law with visual extinctions in the range $ 0 < A_v < 4$. The adopted SFHs are exponentially declining, allowing for e-folding timescales down to 300 Myr.  This choice is simplistic and may not be representative of the real SFHs (e.g., for SFGs see Maraston et al. 2010). However, we are here interested in the stellar mass, which, among SED-derived properties, is most robust against variations (and uncertainties) in model assumptions (e.g., Papovich et al. 2001; Shapley et al. 2005; Maraston et al. 
2010).

In order to conduct further structural analysis using the (binned) output of this SED modeling technique, we constructed pixelized (instead of Voronoi binned) images of the stellar mass surface density distribution in the following way.
First, we construct a $M/L$ map which, within the galaxy's $H_{160}$ segmentation map, has uniform values for pixels belonging to the same Voronoi bins.  Pixels outside the segmentation map (i.e., containing sky noise and possibly faint wings of the galaxy extending below the signal-to-noise threshold) are assigned the average $M/L$ of the nearest three Voronoi bins.  The resulting expanded $M/L$ map is then combined with the \H160 image to construct a final mass map at full $H_{160}$ resolution. This ensures a smooth transition of the galaxy's mass profile from the brighter central parts to the faintest regions.

\subsection{Structural Parameters}
\label{fields_data.sec}

In order to conduct a structural analysis of our galaxy sample, we employ the GALFIT (Peng et al. 2010) morphology fitting code to fit 2-dimensional parametric models to the stellar mass distribution.  For comparison purposes, we model the 2D $H_{160}$ surface brightness distributions as well, following identical procedures.

The models we use comprise on the one hand single \sersic\ models to parametrize the galaxy's shape with the effective radius ($R_e$) and \sersic\ index $n$.  $R_e$ is the galactocentric distance containing half of the total light/mass, and $n$ is a measure of the cuspiness of the overall light/mass profile.  We allow $n$ to vary within the range $0.2<n<8$.  

While the \sersic\ index $n$ is often used and indeed serves as an approximate measure of the contribution of the bulge, it is important to note that $n$ does not translate one to one to the bulge-to-total ratio $B/T$ (see Appendix A; also Andredakis et al. 1995; de Jong et al. 2004; Cibinel et al. 2013; Bruce et al. 2012, Bruce et al. 2014 in prep.).  For a given bulge + disk composite profile, the best-fit \sersic\ index can be increased both by increasing its $B/T$, and by leaving $B/T$ constant while growing the extent of the disk relative to that of the bulge (i.e., lowering $R_{e,B}/R_{e,D}$).  For this reason, and to facilitate a more direct comparison to (semi-analytic) models that are parametrized in terms of bulge and disk units, we furthermore decompose the galaxies by fitting 2-component (bulge + disk) models.  In addition, we also performed one-component pure disk and pure bulge fits by forcing $n=1$ and $n=4$, respectively.

For the bulge-to-disk decomposition, we adopt a procedure similar to that implemented in Bruce et al. (2012) for the bulge+disk decomposition of $H_{160}$ light profiles, where            we fixed the \sersic\ indices to $n=1$ for the disk and to $n=4$ for the bulge.\footnote{Since the choice of a \sersic\ index value for the bulge profile has long been debated in the literature (see Kormendy \& Kennicutt 2004 and references therein), we repeated our two-component fits, with the only change fixing $n=2$ for the bulge component to address the impact on the derived $B/T$ values.  We find a median difference of $B_{n=4}/T - B_{n=2}/T = 0.03^{+0.15}_{-0.11}$ for quiescent galaxies and $0.01^{+ 0.09}_{-0.12}$ for SFGs, with the errors marking the 1$\sigma$ scatter.  Any systematic trends are small, and our results are therefore robust against the precise value of $n_{bulge}$ adopted.}  The centers are left free, but we restrict the relative distance between the bulge and disk centers to be less than 2 pixels.  
All other parameters defining the two components ($R_{e}$, the axial ratio $b/a$, and the total magnitude/mass of both components) are allowed to vary independently.  

In 2-component modeling, the higher number of degrees of freedom increases the odds of the fit being trapped in a local \chis\ minimum.  In order to mitigate this risk, we ran GALFIT using a grid of initial starting values.  Our grid was constructed by using a range of size ratios between bulge and disk ($R_{e,B}/R_{e,D}$) ranging from 0.1 to 1, in steps of 0.1.  For each initial guess of $R_{e,B}/R_{e,D}$, the corresponding initial guess on $B/T$ was then set such that the \sersic\ index matching this initial configuration matches the one measured in the single-component fit (see Appendix A).  Likewise, the initial magnitudes and absolute values of the initial size guesses for bulge and disk were set such that the total magnitude and half-light/mass radius of the composite profile matches the respective values determined from \sersic\ fitting.  

In cases where GALFIT yields solutions with implausibly small bulge sizes ($< 0.1$ pixel, corresponding to $\lesssim 1/30$ of the resolution, or $\lesssim$ 50 pc at $z\sim2$) and flags the outcome as potentially not converged and unphysical, we excluded the respective run from the grid.  Also, solutions yielding a disk smaller in size than the bulge (i.e., $R_{e,B}/R_{e,D} > 1$) were not included.\footnote{We note that nuclear stellar disks in early-type galaxies do exist (e.g., Jaffe et al. 1994; van den Bosch et al. 1994; Ferrarese et al. 1994), but they are impossible to resolve at the HST resolution for $z \sim 2$ galaxies.}  

After performing the fits for each point of the grid with initial guesses, we assigned a final $B/T$ ratio for each object as the solution of the fit with the lowest $\chi_{red}^2$.  Here, the pure disk and pure bulge fits were also included.  Their solutions generally show higher $\chi_{red}^2$ than the 2-component fits, but are occasionally preferred over those in a $\chi_{red}^2$ sense (oftentimes, these are bulgeless systems with $n < 1$).  
The two-component decompositions are statistically preferred over the single \sersic\ models (as based on both their $\chi_{red}^2$ values and the Akaike information criterion, AIC\footnote{In evaluating models with the AIC, the preferred model is the one that minimizes $\chi^2 + 2p + \frac{2p(p+1)}{N-p-1}$, where $p$ is the number of free parameters in the fit and $N$ is the number of data points used in the fit.}) for $\sim2/3$ of the total sample.  Those systems for which the single \sersic\ model yields a lower value of $\chi_{red}^2$ and AIC typically feature shallow profiles with  $n < 1$.

In both single- and two-component fitting, we use an automated scheme which pre-determines neighboring sources that need to be masked or fitted simultaneously, and passes initial guesses of fitting parameters to GALFIT.  Those include estimates on size, total magnitude and center.  The initial guess on size is based on the distance from the center to the radius at which the curve of growth reaches 50\% of the galaxy's total flux.  The mass-weighted center of the galaxy derived within its segmentation map is adopted as initial estimate for the center.  
GALFIT takes into account the convolution of the model with the point-spread-function (PSF).  Both for fitting light and mass, we use a PSF which is a combination of stacked stars and a Tiny-Tim \citep{krist1995} model PSF.  For a more detailed description of the used PSF, see \cite{vdw2012}.

We emphasize that throughout this paper our working definition adopted for bulge and disk components is based on the above bulge+disk modeling of stellar mass or light maps, as empirical constraints on whether or not stars assigned to a bulge/disk component are dynamically hot/cold are currently lacking.  As the SINS survey of $z\sim2$ galaxy kinematics demonstrated that high-redshift SFGs are dynamically (F\"{o}rster Schreiber et al. 2009) and morphologically (F\"{o}rster Schreiber et al. 2011) distinct from local spiral galaxies, we caution that bulge fractions as derived by our decompositions may have a somewhat different meaning than they would have in the local universe. 

In addition, we estimated a typical measurement uncertainty on $B/T$ by setting up an array of model galaxies with a range of mass, size, $B/T$ ratio, and $R_{e,B}/R_{e,D}$, which is similar to our data.  The grid of model galaxies consisted of 5, 3, 11, and 10 grid points in mass, $R_e$, $B/T$ and $R_{e,B}/R_{e,D}$ respectively.
These were then inserted in multiple empty sky regions of the CANDELS UDS field to mimic the typical background noise\footnote{UDS is part of CANDELS-Wide, which was exposed for two HST orbits divided over F125W and F160W.  Part of our data set comes from the CANDELS-Deep regions (the central halves of the GOODS fields), which received 4 orbits per pointing in F125W and F160W each.  The inferred uncertainties from our analysis of mock galaxies can therefore be considered as conservative estimates.}.  We next ran GALFIT on all 8250 mock galaxies using our 2-component fitting scheme.  The measurement error on $B/T$ of each galaxy in our sample given its mass, radius, and profile shape is then finally assigned as the scatter among the recovered $B/T$ ratios of the corresponding model galaxy.

Typical measurement errors in $B/T$ for star-forming and quiescent galaxies are on average $\sim0.05$ and $\sim0.06$ at $z\sim 1$ and $\sim0.1$ and $\sim0.13$ at $z\sim 2$, respectively.  The distribution of errors peaks below the median ($\lesssim0.05$), and shows a tail towards higher $B/T$ errors.  Two alternative methods to estimate the uncertainty in $B/T$, namely the formal random uncertainty reported by GALFIT and re-fitting the observed galaxies after applying additional background and Poisson noise, generally lead to lower estimated uncertainties (by a factor of $ \sim 2$ in the case of GALFIT).  In the remainder of the paper, we therefore adopt the most conservative error estimates inferred from our analysis of the inserted mock galaxies.

\section {Results on Galaxy Structure}
\label{structure.sec}

\subsection{The Evolving Mass Budget of Disks and Bulges}
\label{budget.sec}

Exploiting the bulge-disk decompositions of the stellar mass maps derived for our sample of massive galaxies, we first evaluate the average mass budget of disks and bulges.  Let us consider picking a random star  out of our sample of massive galaxies above $10^{10}\ M_{\sun}$.  At $1.5<z<2.5$, the probability that this star belongs to a bulge component is $46\%$.  Increasing the mass limit to $\log(M) = 10.5$ or 11 yields a higher probability for the star to be associated to the bulge, of $49\%$ and $54\%$, respectively.  
Perhaps somewhat surprisingly, the fraction of stars residing in a bulge component rises only slightly to $0.5<z<1.5$, to $47\%$, $50\%$, and $56\%$ for galaxies more massive than $\log(M) = 10$, 10.5, and 11 respectively.  

The formal uncertainties to the above stated probabilities including sample variance and typical measurement errors on $B/T$ are limited to a few percent.  The total error budget is likely dominated by systematics, for example related to the assumptions made in stellar population modeling (see Section 3.1).  We note, however, that only $M/L$ uncertainties with a differential impact on bulges and disks will affect the above numbers.  Even if the $M/L$ ratio of bulges were systematically under- or overestimated by 0.2 (0.3) dex with respect to those of disks, the change in the above numbers would be limited to $\sim 7 (10)\%$.

As bulges, unlike stellar disks, can be considered sinks in the continuous assembly of a galaxy's stellar component, the rising mass density of stars in bulges (by a factor of $\sim 1.8$ from the higher to the lower redshift bin) therefore seems to be compensated largely by the continuing assembly of new disks.  Splitting our sample in finer redshift bins, we do find the fraction of stellar mass in bulges to increase more significantly, by a factor $\sim 1.5$ over the entire 6 Gyr timespan sampled by our study.
 
Overall, the bulge mass fractions are higher than what would be inferred from fits to the $H$-band surface brightness profiles, as the median mass-to-light ratio of disk components is 0.2 dex lower than of bulge components. 
The above numbers address the evolving mass budget of disks and bulges for a mass-limited sample including both star-forming and quiescent galaxies.  In the remainder of the paper, we will delve into more depth by breaking down our sample by star formation activity.

 \begin {figure*}[htbp]
 \centering
   \includegraphics[width=\textwidth]{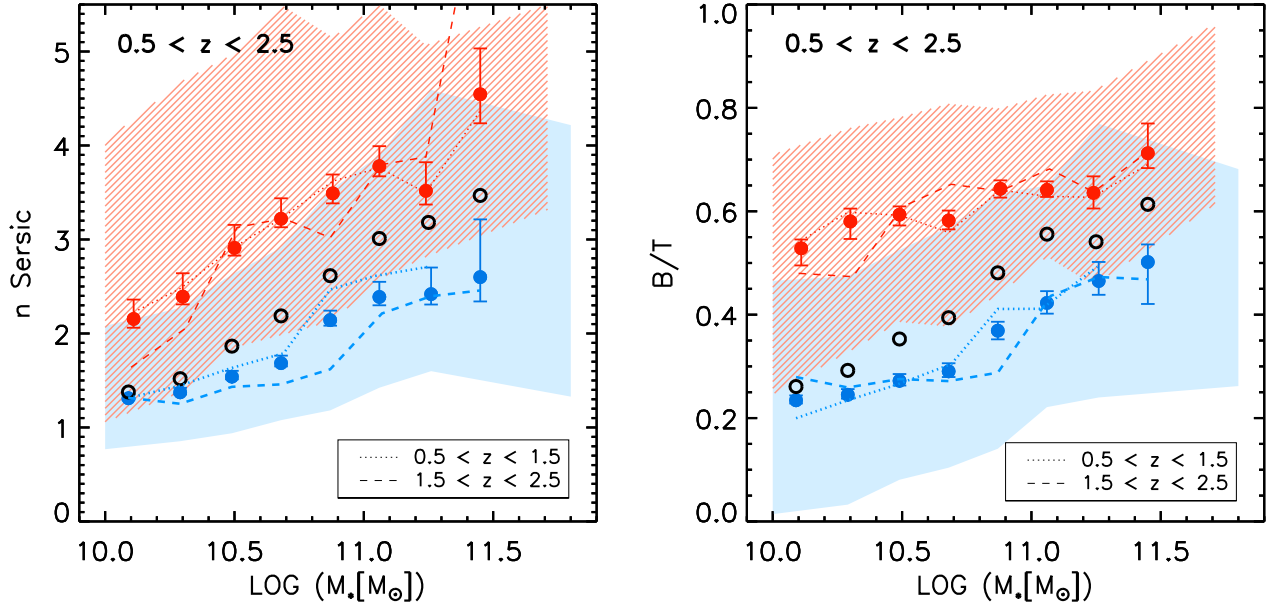}
 \caption{
 \sersic\ index and $B/T$ ratio as a function of total stellar mass of our galaxy sample spanning the redshift range $0.5 < z < 2.5$. The whole sample (black non-filled symbols) is divided in star-forming (blue symbols) and quiescent galaxies (red symbols).  The respective shaded areas mark the 50th percentile scatter of the distributions within one bin, while the error bars indicate the uncertainty on the median value.  Results for the lower and upper halves of the redshift interval are marked with dotted and dashed lines, respectively.  The bins contain (from left to right) 1439, 1091, 886, 693, 484, 259, 114, 35 star-forming systems and 227, 267, 298, 369, 280, 214, 83, 25 quiescent systems within the whole redshift range.  SFGs show a clear trend of increasing \sersic\ index and $B/T$ with increasing stellar mass, reaching $B/T \sim 0.4-0.5$ above $10^{11}\ M_{\sun}$, reflecting the build-up of central mass concentrations in main-sequence SFG up to $z \sim 2.5$.
 }
 \label{struc.fig}
 \end {figure*}

\subsection {Profile shape}
\label{profile.sec}

In recent years, several HST-based studies have investigated the structural differences between star-forming and quiescent galaxies at high redshift.  In common to all of these analyses, star-forming systems are found to have significantly larger rest-optical sizes than their quiescent counterparts at the same mass and redshift (e.g., Toft et al. 2009; van der Wel et al. 2014).  In addition, their surface brightness profile shapes tend to be shallow ($n \sim 1$), while quiescent galaxies feature cuspier light profiles (e.g., Wuyts et al. 2011; Bell et al. 2012; Cheung et al. 2012, although see also Bruce et al. 2012 for decomposed light profiles at high stellar masses and redshifts).

At the same time, the same multi-wavelength high-resolution lookback surveys have also established that substantial mass-to-light ratio variations in the rest-optical can occur, not only between but also within galaxies (see, e.g., Wuyts et al. 2012; Guo et al. 2012; Boada et al. in prep).  Typically, as new stars tend to form from gas settled in a disk configuration, such $M/L$ ratio variations are anticipated to give rise to a composite light profile in which the disk component has a relatively larger weight (per unit mass) than the bulge.   It is therefore important to address to which degree the above structural distinction between the two classes of galaxies is intrinsic to their distribution of stellar mass, or, conversely, can be attributed to stellar population effects.  The answer to this question is of immediate relevance to our understanding of quenching, as in principle the latter scenario could imply that compact quiescent systems can evolve from the star-forming main sequence by simple fading, 
without invoking an associated morphological transition.  Kriek et al. (2009) investigate this scenario for a spectroscopically confirmed sample of massive $z \sim 2$ galaxies, finding that 3 out of 6 massive star-forming systems have dense cores, and thus may passively evolve into compact galaxies due to fading of the outer star-forming regions.  Szomoru et al. (2010), on the other hand, exploit the exquisite depth of the Hubble Ultra Deep Field to probe the surface brightness profile of a massive compact quiescent galaxy at $z = 1.91$, ruling out the existence of a faint extended envelope or disk around the observed galaxy.  Another argument against fading comes from Cheung et al. (2012), who derived the stellar masses of bulges for both star-forming and quiescent galaxies at $0.5 \leqslant z < 0.8$. They found that bulges of SFGs are half as massive as those of similar-mass quiescent galaxies, implying they cannot simply fade onto the red sequence without structural evolution.  

Using the stellar mass maps reconstructed for our mass-selected sample of 6764 galaxies  at $0.5 < z < 2.5$ with $\log(M) > 10$, we are now able to draw statistically significant conclusions on the structural distinction between high-z galaxies prior to and after quenching.  In Figure\ \ref{struc.fig}, we compare the shape of the stellar mass distributions (i.e., corrected for spatial $M/L$ variations) of star-forming and quiescent galaxies, and study their dependence on the total galaxy stellar mass.  We consider profile parameters based on single-component (i.e., \sersic ) fits as well as two-component (bulge + disk) decompositions, and show the results for two separate redshift intervals: $0.5 < z < 1.5$ and $1.5 < z < 2.5$.  In both cases, we identified a galaxy as quiescent if its specific SFR (sSFR) satisfied $sSFR < \frac{1}{3 t_{Hubble}}$, where $t_{Hubble}$ is the Hubble time at the redshift of the galaxy, and as star-forming otherwise.  We tested that a definition of quiescence based on the 
location of a 
galaxy in the UVJ diagnostic diagram (Wuyts et al. 2007; Williams et al. 2009) yields effectively identical results.

Figure\ \ref{struc.fig} immediately highlights that the distinct structural appearance of star-forming and quiescent galaxies is intrinsic to its internal distribution of stellar mass, and not just driven by stellar population or obscuration effects.  In fact, a comparison to the equivalent plots based on $H$-band surface brightness profiles rather than mass maps (see Appendix B) indicates that stellar population effects (when measuring at rest-optical wavelengths) only induce a modest, albeit non-negligible shift.  At all masses, quiescent galaxies feature cuspier stellar mass distributions (i.e., higher $n$) than star-forming systems.  Their typical best-fit \sersic\ index is furthermore an increasing function of galaxy mass.  Interestingly, also among SFGs the profile shape is not independent of stellar mass.  An increase in $n$ is apparent above $10^{11}\ M_{\sun}$, both at $z \sim 1$ and at $z \sim 2$.  A similar trend of increasing cuspiness at the tip of the MS was noted by Wuyts et 
al. (2011, Figure 1; see also Nelson et al. in prep).

Next, it is worthwhile reflecting on what it is that we measure when fitting \sersic\ profiles.  Appendix A illustrates that, when considering galaxies as superpositions of bulge and disk components, a given best-fit \sersic\ index does not necessarily correspond one-to-one to a unique $B/T$ value, even though it is often interpreted as such.  Given a bulge+disk system with associated best-fit $n$, one can increase its $n$ either by boosting $B/T$, or, alternatively, by growing the extent of the disk with respect to that of the bulge without any change to $B/T$.

Turning to the right-hand panel of Figure\ \ref{struc.fig}, we now explore the $B/T$ ratio as a function of galaxy mass, for SFGs and quiescent galaxies separately.  Again, we find a clear anti-correlation between star formation activity and bulge prominence.  Focussing on the star-forming population, the median $B/T$ is limited to below 30\% for intermediate mass SFGs ($10 < \log(M) < 11$), while typical bulge mass fractions rise to 40-50\% above $10^{11}\ M_{\sun}$.  We note that there is a significant scatter in the distribution of individual $B/T$ values around the median for both quiescent and star-forming galaxies.  We investigated the variation in median trends when varying the binning intervals, finding negligible changes at lower masses, while the median $B/T$ of the most massive ($\log(M) \sim 11.3$) SFG bin changes by $\pm 0.1$, depending on the applied binning intervals\footnote{The binning scheme applied in Figure\ \ref{struc.fig} is such that the most massive bin still contains more than 10 
galaxies, allowing a robust estimation of the median.}.

We note that measurements on the $H$-band yield bulge mass fractions among SFGs that are lower by on average $\sim 30\%$, as can be understood from a disk component composed of a younger, lower $M/L$ stellar population than the bulge.

 \begin {figure*}[htbp]
 \centering
   \includegraphics[width=\textwidth]{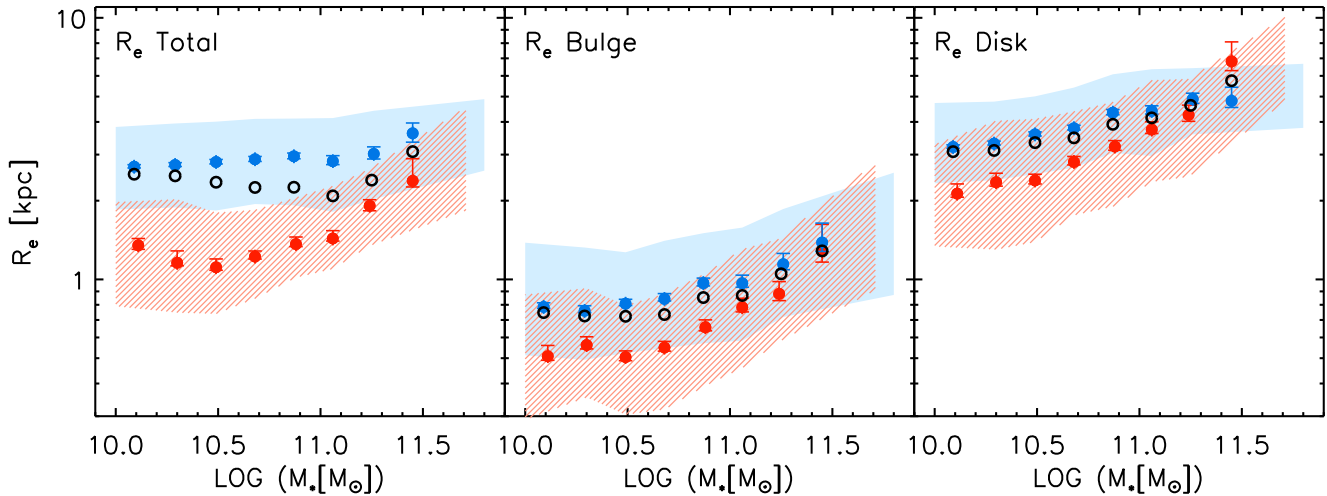}
 \caption{Effective total size, bulge size and disk size as a function of total stellar mass in the redshift range $0.5 < z < 2.5$. The total sizes are derived from single \sersic\
 fits, bulge and disk sizes are inferred from the bulge-disk decompositions.  The whole sample (black symbols) is divided into star-forming (blue symbols) and quiescent galaxies (red symbols).  The respective shade areas mark the 50th percentile scatter of the distributions, while the error bars indicate the uncertainty on the median value.  There is a clear difference in total size between SFGs and QGs over the whole mass range.  Also the bulge and disk components individually appear to be smaller for the star-forming population up to masses of $\log(M) \sim 11.3$.
}
 \label{Radii.fig}
 \end {figure*}

From the two-component fits, we infer a typical $R_{e,B}/R_{e,D}$ size ratio of $\sim 0.2$, albeit with significant scatter (see Figure\ \ref{n_bd.fig}).  The median size ratio shows little dependence on star formation activity or mass, over the range probed by our sample.  
Given the enhanced $B/T$ values in quiescent galaxies, and the fact that bulges have smaller half-mass radii than disks, one may wonder if the difference in total size between SFGs and quiescent galaxies can be accounted for completely by a redistribution of stellar material from the disk to the bulge, without changing the extent of each of the components individually.  Our analysis confirms that the change in $B/T$ of SFGs prior or during quenching is to a large extent responsible for the size difference between the quiescent and star-forming population.  However, some fraction of the shrinking size is still attributed to the individual components being smaller.  
In Figure\ \ref{Radii.fig}, we show the total sizes as well as the sizes of the individual components for star-forming and quiescent galaxies, as measured on the mass maps.  While the total sizes of SFGs and QGs are noticeably different (by a factor $\sim 3$ at $\log(M_*) \sim 10.5$), the difference in size of bulge and disk components between SFGs and QGs respectively is smaller, typically by a factor $\sim 1.5$.

\begin {figure*}[htbp]
\centering
  \includegraphics[width=0.8\textwidth]{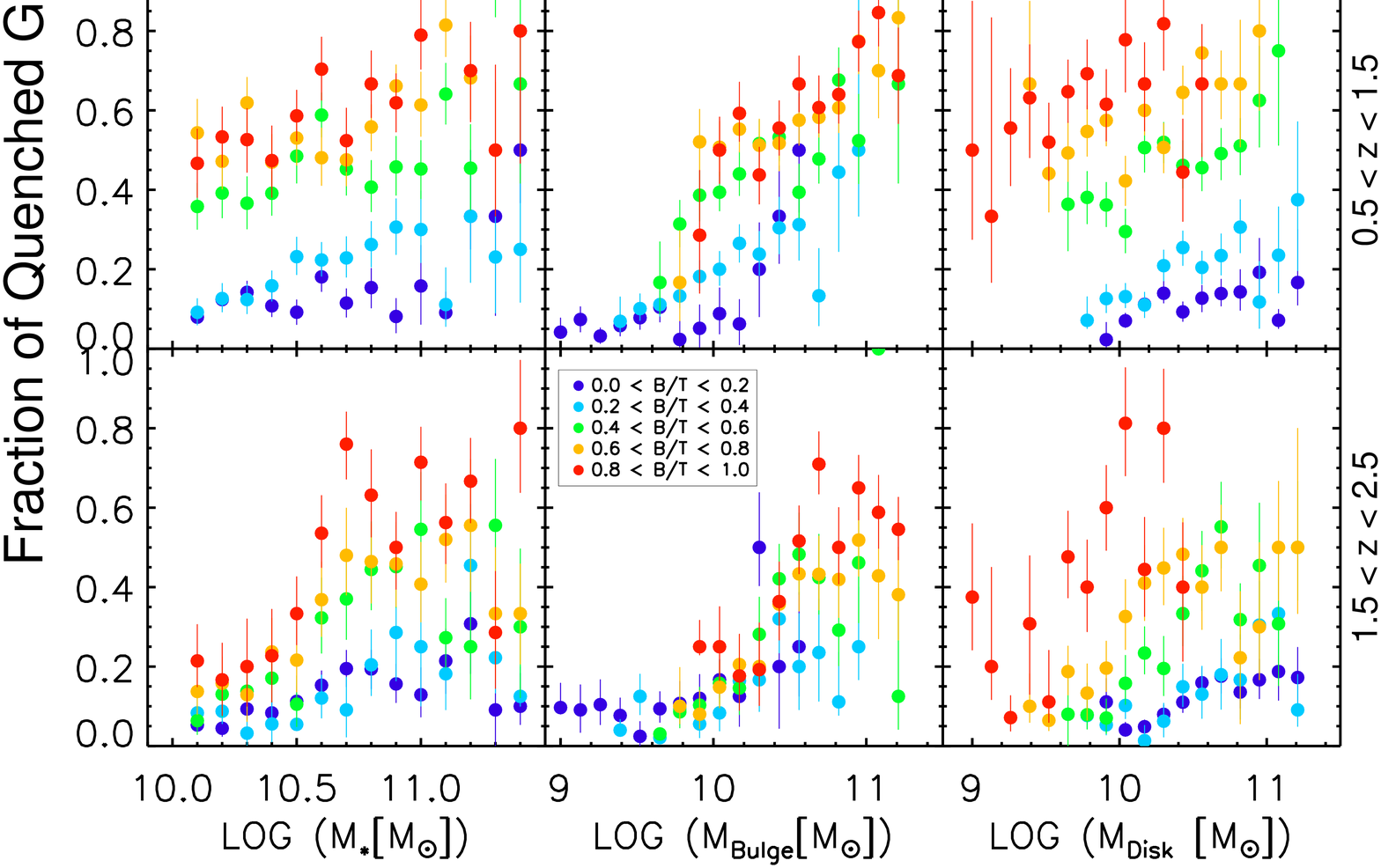}
\caption{
Fraction of massive galaxies ($M > 10^{10}\ M_{\sun} $) that are quiescent ($f_{quench}$) as a function of total stellar mass (left columns), bulge mass (middle columns) and disk mass (right columns) for $z \sim 1$ and $z \sim 2$. Galaxies with $sSFR < 1/3*t_{Hubble}$ are assigned as quiescent, the others as star-forming.  In the middle and bottom rows, we split the galaxy sample in bins of $B/T$ for the two redshift bins.  Uncertainties are derived from bootstrapping, and include sample variance as well as measurement uncertainties.  
A positive trend of $f_{quench}$ is seen with total stellar mass, with this trend becoming stronger when correlating $f_{quench}$ with the bulge mass, whereas $f_{quench}$ shows no positive correlation with the mass of the disk component.  At a given galaxy mass, $f_{quench}$ is increasing significantly with increasing $B/T$ ratio.  The scatter in $f_{quench}$ among the different $B/T$ bins is largely reduced when correlating against the bulge mass, implying that the mass within a bulge of a galaxy is correlating best with quiescence.  This trend is qualitatively similar for both redshift ranges, with overall quenched fractions being lower at $z \sim 2$ than at $z \sim 1$.
}
\label{fquench_all.fig}
\end {figure*}

\subsection {Fraction of Quenched galaxies}
\label{f_quench.sec}

With the morphological parameters of the mass maps for our entire galaxy sample in hand, we now proceed to relate those with galaxy-integrated star formation properties. 

The three panels of Figure\ \ref{fquench_all.fig} show, from left to right, the fraction $f_{quench}$ of quenched galaxies as a function of total stellar mass, bulge mass, and disk mass, respectively.  Here, we again define galaxies as quenched/quiescent when $sSFR < 1/3*t_{Hubble}(z)$, and as star-forming otherwise.  

The uncertainties in $f_{quench}$ are derived via a bootstrapping method and represent the 68\% confidence levels.  They include both sample variance and the typical measurement errors on $B/T$, which are derived as described in Section 3.2.  For the bootstrapping, we computed $f_{quench}$ for 1000 samples, which are randomly drawn from the original sample, with replacement.  For each bootstrap iteration, we displace the $B/T$ values for each galaxy by the typical measurement error in $B/T$, given the galaxy's magnitude, size and measured profile shape.

The top left panel of Figure\ \ref{fquench_all.fig} illustrates that the fraction of quenched galaxies increases with increasing mass, from $\sim 0.1$ at around $10^{10} M_{\odot}$ to $\sim 0.5$ at $2 \times 10^{11} M_{\odot}$.  
A second conclusion is that the fraction of quenched galaxies is overall higher for the lower redshift bin, by a factor of $\sim2$ on average.  Both of these results are well established in the literature.  The rising mass function of quiescent galaxies over cosmic time has most recently been quantified on a firm statistical footing by Muzzin et al. (2013) and Ilbert et al. (2013), both of which exploit the wide-area UltraVISTA survey.  What CANDELS lacks in number statistics compared to UltraVISTA, it adds in depth and high resolution.  Exploiting these key strengths, we now turn to the dependence of the quenched fraction on galaxy subcomponents: the mass of their bulge (middle panel) and disk (right-hand panel).  Clearly, the dependence of  $f_{quench}$ on the bulge mass is much stronger than on the disk mass, which does not show any significant correlation with $f_{quench}$ above $\log(M_{Disk}) \sim 9.5$.  Towards lower disk masses, $f_{quench}$ increases rapidly, but we point out that this trend is 
entirely driven by the (total) stellar mass limit of our sample ($\log(M_*) > 10$; i.e., the galaxies occupying the lowest $M_{Disk}$ bins are necessarily heavily bulge-dominated systems, that tend to form relatively few stars).  If lower mass galaxies were to be included, less massive, disk-dominated SFGs would likely outnumber these massive spheroids with small residual disks in the low $M_{Disk}$ bins, producing a flat relation of $f_{quench}$ with $M_{Disk}$ over the full range probed.  
Above respective masses of $10^{10}\ M_{\sun} $, $f_{quench}$ increases more rapidly with bulge mass than with total stellar mass in both redshift ranges ($\sim 0.35$ per dex of $M_{Bulge}$ compared to $\sim 0.3$ per dex $M_*$, or $\sim 0.1$ per dex of $M_{Disk}$).

With the bulge-to-disk decompositions in hand, we next split the galaxy sample in bins of $B/T$, and explore second parameter dependencies.  Considering first the dependence of $f_{quench}$ on the total stellar mass, it is apparent that, at a given total mass, $f_{quench}$ is increasing significantly with increasing $B/T$ ratio.  The middle panels of Figure\ \ref{fquench_all.fig} illustrate that, when considering the dependence of $f_{quench}$ on bulge mass, the different $B/T$ bins align along a much tighter locus.  In contrast, a large spread is seen as a function of disk mass (right-hand panels of Figure\ \ref{fquench_all.fig}).  
In order to quantify these trends, we compute the Spearman's rank correlation coefficient ($r_{s}$) for the relations of $f_{quench}$ with $\log(M_*)$, $\log(M_{Disk})$, and $\log(M_{Bulge})$ for respective masses $\log(M) > 10$.  We find that $r_s$ is indeed significantly higher for the relation $f_{quench}$ vs. $\log(M_{Bulge})$ ($r_{s} \sim 0.68$) than for both $f_{quench}$ vs. $\log(M_{*})$ ($r_{s} \sim 0.32$) and $f_{quench}$ vs. $\log(M_{Disk})$ ($r_{s} \sim -0.05$), as measured for $z \sim 1$.  Consistent results are found for $z \sim 2$. 

We investigated the impact of defining quiescence based on a $UVJ$ color-color criterion instead of a sSFR cut.  When applying a UVJ-based selection of quiescent galaxies, we find an overall good agreement with the trends presented in Figure\ \ref{fquench_all.fig}.   Quantitatively, small changes occur, with $f_{quench}$ increasing by $ \sim 7$\% for the entire $0.5 < z < 2.5$ sample integrated over all masses.  The good agreement is not surprising, since the precise threshold in sSFR used to select quiescent galaxies ($sSFR < 1/3*t_{Hubble}(z)$) was chosen to yield maximum overlap with the $UVJ$ selection criterion.

Taken together, this demonstrates that the build-up of a bulge seems to play a critical role in the quenching process of galaxies, whereas the disk does not.  The amount of stars in the disk component of a galaxy has little to no predictive power regarding its star-forming or quenched state, unless also $B/T$ (and hence the bulge mass) is known.  We find a qualitatively similar behavior at $z \sim 2$ as at $z \sim 1$, but note that the cosmic evolution in the quiescent fraction cannot solely be attributed to continuing bulge growth over time, as galaxies in the same $M_*$ and $B/T$ bin at $z \sim 1$ are more likely to be quenched than those at $z \sim 2$.  Appendix B illustrates how the equivalent diagrams composed from fits to the $H$-band surface brightness rather than the stellar mass distribution exhibit a larger spread from low to high $B/T$ bins.  This generic behavior can be understood from a physical picture where the disk component has a relatively larger weight in light than in mass.

Our work is in agreement with, and takes the next step beyond previous reports that the inner stellar mass density is better related to the star formation history than the total stellar mass \citep{franx2008,bell2012}, as inferred from rest-optical imaging of smaller samples of high-redshift galaxies (see Kauffmann et al. 2003 and Fang et al. 2013 for a local universe reference, and Cheung et al. 2012 for intermediate redshifts $z<0.8$)

Importantly, the same behavior explored here over the redshift range $0.5 < z < 2.5$ extends in a strikingly similar fashion all the way to the present day, as demonstrated by Bluck et al. (2014) who exploit the large number statistics of SDSS.

\section {Comparison with SAMs}
\label{SAM_results.sec}

\subsection{The Somerville model}
\label{somerville.sec}

Semi-analytic models (SAMs) have a rich history of trying to reproduce galaxy scaling relations and abundances, with the goal of guiding our interpretation of the observational results.  Here, we focus specifically on the SAM developed by Somerville et al. (2008) and further updated by Somerville et al. (2012) and Porter et al. (2014), which is rooted in the Bolshoi cosmological dark matter simulation (Klypin et al. 2011).\footnote{Hereafter, we refer to this model as the Somerville et al. SAM.}  As is generic to all SAMs, the model relies on simplified analytic prescriptions for the dynamical and astrophysical processes down from entire galaxy scales, rather than on kiloparsec to parsec scales (the resolution below which state-of-the-art cosmological and zoom-in hydro-simulations resort to subgrid physics, respectively).  This limitation, however, yields the enhanced flexibility of a relatively inexpensive runtime, allowing the straightforward generation of statistically significant model galaxy populations,
 and the tuning of parameters to observational constraints such as mass functions and scaling relations (only empirical constraints from the nearby universe 
were used in tuning the parameters of the model considered here).  The fact that SAMs conceptually are formulated in units of bulge and disk components furthermore makes them suitable for a direct and meaningful comparison to the diagnostics explored in this paper.

A detailed description of the prescriptions for cooling, star formation, feedback, and structural growth is provided by Somerville et al. (2008, 2012), with extensions and applications to the CANDELS data set presented by Porter et al. (2014).  The input to and output from the model is further contrasted to that of other SAMs by Lu et al. (2013).  

For a detailed discussion of the physical recipes of this SAM and the resulting output in the context of a larger set of SAMs, we refer the reader to Lu et al. (2013).  
For the sake of the comparison presented here, we emphasize that none of the relations investigated in this paper formed part of the set of observational constraints to which the model parameters were tuned.  Model parameters of the SAM were tuned to (approximately) match the global stellar mass function, the stellar mass function of early- and late-type galaxies, the gas fraction as a function of stellar mass for disks, and the mass-metallicity relation for stars.  

Also of particular relevance is the fact that, in the model, stars form either in the disk following a Kennicutt-Schmidt law (Kennicutt 1998), where the disk scalelength is set by a similar methodology as Mo, Mao \& White (1998), or, in the event of a merger, during a starburst.  Bulge formation as well as feeding of the central supermassive black hole can happen through two channels: mergers or disk instabilities (see Porter et al. 2014).  The starburst, black hole accretion and morphological transformation induced by mergers depends on the mass ratio and gas fraction, as calibrated using a large suite of binary merger simulations (Somerville et al. 2008; Hopkins et al. 2009; Somerville et al. 2012).  Star formation is moderated through heating of gas by supernovae as well as through AGN feedback.  In addition to the quasar mode, during which AGN can drive powerful outflows, black holes also grow more gradually over longer timespans through the so-called radio mode (i.e., suppression of cooling via radio 
jets).  No explicit connection between the 
bulge mass and quenching (as may for example be expected from the Toomre Q stability criterion in a gravitational quenching scenario, see Section 6.2) was built into the model.

\begin {figure}[t]
\centering
 \includegraphics[width=0.49\textwidth]{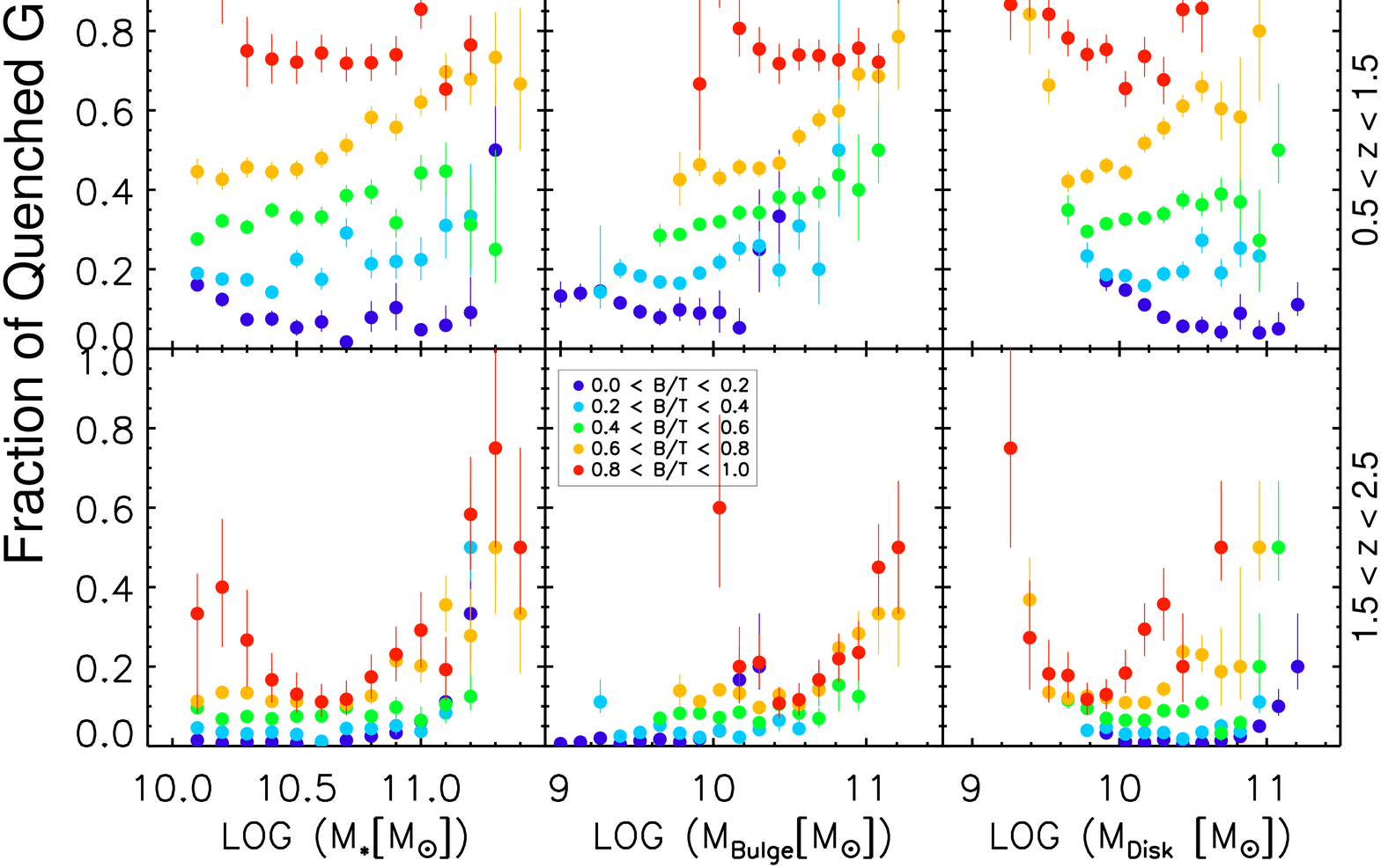}
\caption{Same as Figure\ \ref{fquench_all.fig} using galaxies from the Somerville et al. SAM.  Galaxies with $sSFR > 1/3*t_{Hubble}$ are assigned as star-forming, the others as quiescent.  Uncertainties are derived via bootstrapping to reflect the sample variance. 
\label{SAM_Somerville1.fig}}
\end {figure}

\subsection{$f_{quench}$ in the SAM}
\label{fquench_SAM.sec}

In Figure\ \ref{SAM_Somerville1.fig}, we show the model equivalent of Figure\ \ref{fquench_all.fig}, describing how the fraction of galaxies that are quenched depends on the total stellar mass, the mass of stars in the bulge, and the mass of stars in the disk component respectively.  As for the observations, we define the threshold for a galaxy to be quenched with a ruler moving with redshift: $sSFR < 1 / (3 t_{Hubble})$.  This definition, rather than an application of the UVJ diagnostic, remains closer to the direct output of the SAM, avoiding a translation to mock spectral energy distributions, which would introduce additional assumptions and uncertainties \footnote{It should be noted that equivalent assumptions and uncertainties associated with the conversion from light to physical properties enters upon SED modeling of the observed galaxies.  The choice of how far to take the models to the observations or visa versa therefore remains somewhat arbitrary.}.

At first glance, the SAM features several of the characteristic trends noted earlier for the CANDELS galaxies: $f_{quench}$ rises towards later cosmic times, increases with the total stellar mass, more steeply so with bulge mass, and shows no appreciable correlation with disk mass above $\log(M_{Disk}) = 9.5$.  In the interval $9 \lesssim \log(M_{Disk}) \lesssim 9.5$, a sharp drop with increasing $M_{Disk}$ is noted, as also seen in the observations.  Given the $M > 10^{10}\ M_{\sun}$ threshold of our sample selection, the latter objects are necessarily heavily bulge-dominated.  Without imposing such a mass limit, less massive, disk-dominated SFGs would outnumber massive early-type galaxies with small residual disks in the lower $M_{Disk}$ bins.  Despite the qualitative success, quantitative differences in the quenched fractions of model galaxies with respect to those observed are clearly present.  The discrepancy is most severe at $z \sim 2$, where modeled $f_{quench}$ values are, on an average over the 
whole displayed mass range, of order a factor $\sim 2 - 3$ short of observed, hinting at an underestimated quenching rate and/or inefficiency to prevent quiescent systems from rejuvenating\footnote{See also Ciambur et al. 2013 for a discussion on the quenched fraction in the Garching semi-analytic models.}.

When splitting the SAM galaxies at each $M_*$, $M_{Bulge}$, and $M_{Disk}$ in bins of $B/T$ (middle and bottom panels of Figure\ \ref{SAM_Somerville1.fig}), we reproduce a similar behavior as found for the real universe in Section\ \ref{f_quench.sec}.  Namely, the total stellar mass acts as a poorer predictor of the quenched state of a galaxy. 
This situation can be remedied if in addition to $M_*$ also $B/T$ (and hence $M_{Bulge}$) is known.  Quantitatively, the correlation between $f_{quench}$ and $M_{Bulge}$ is measured to be the strongest ($r_s \sim 0.46$ ), whereas $M_*$ and $M_{Disk}$ only show weak correlation with $f_{quench}$ ($r_s \sim 0.21 $ and $r_s \sim -0.1$, respectively). The quoted values of $r_s$ are measured for respective masses of $\log(M) > 10 $ and at $z \sim 1$.  The values for $r_{s}$ at $z \sim 2$ are similar, with the correlation between $f_{quench}$ and $\log(M_{*})$ as well as between $f_{quench}$ and $\log(M_{Disk})$ being somewhat stronger.  
At $0.5 < z < 1.5$, less than 20\% of all massive galaxies with $B/T < 0.2$ are classified as quiescent.  Conversely, the majority of galaxies in the upper $B/T$ bin (with $B/T > 0.8$) have low sSFR.  These inferences are in common between the SAM and the observations.  Also in agreement, is the fact that $M_{Bulge}$ serves as a better predictor of $f_{quench}$ than the total stellar mass, with different $B/T$ bins being more (albeit not perfectly) aligned along a single locus in the $f_{quench}$ versus $M_{Bulge}$ diagram.

At $1.5 < z < 2.5$, the model predictions are skewed towards too low $f_{quench}$ values, as noted earlier.  However, in relative terms the same generic behavior as a function of bulge prominence is notable.

We note that most of the trends in Figure\ \ref{SAM_Somerville1.fig} are driven by physical prescriptions in the SAM affecting central galaxies rather than satellites, as centrals account for 80\% (90\%) of the model galaxy population above $log(M) = 10$ (11).  Those massive galaxies classified as satellites are further subjected to additional environmental quenching processes, resulting in a higher $f_{quench}(M_*,M_{Bulge})$ for this particular subpopulation.

\subsection{The agent of quenching}
\label{agent_SAM.sec}

\begin {figure}[htp]
\vspace{-5mm}
 \centering
  \includegraphics[width=0.45\textwidth]{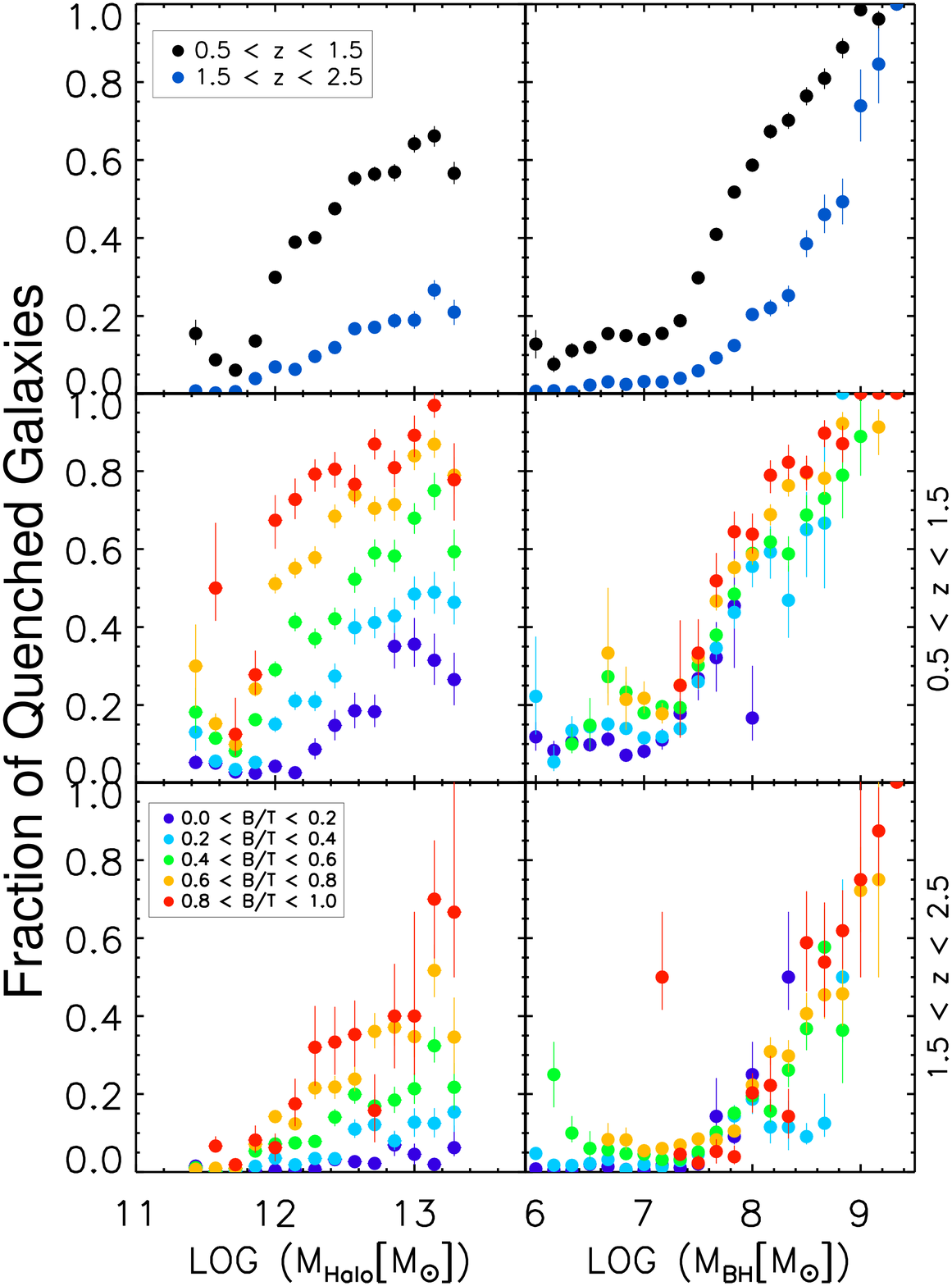}
\vspace{-2mm}
\caption{$f_{quench}$ as a function of halo and black hole mass using galaxies from the Somerville et al. SAM.  In the middle and bottom rows, the galaxy sample has been split in bins of $B/T$.  Galaxies with $sSFR > 1/3*t_{Hubble}$ are assigned as star-forming, the others as quiescent.  Uncertainties are derived via bootstrapping to reflect the sample variance.    
\label{SAM_Somerville2.fig}}
\end {figure}

Given the qualitative agreement between model and observations, we can now pose the question how, in the context of the Somerville SAM, the  relation between structure and stellar populations could be interpreted.  To this end, we consider the dependence of $f_{quench}$ on two physical properties of galaxies in the SAM that are not observationally accessible for our CANDELS sample: the halo mass $M_{Halo}$ and the mass of the central supermassive black hole $M_{BH}$.  The top panels of Figure\ \ref{SAM_Somerville2.fig} illustrate that the probability of a galaxy being quenched increases towards high $M_{Halo}$ and high $M_{BH}$.  In detail, however, the dependencies on the two look different.  The increase in $f_{quench}$ with $M_{Halo}$ is gradual over nearly two orders of magnitude.  In contrast, a much sharper upturn of $f_{quench}$ emerges above $\log(M_{BH}) = 7.5$.   This behavior is especially notable at $z \sim 1$, but a rise above the same threshold is present in the $z \sim 2$ population as well.   
Breaking the model galaxy population down by its structural properties, we find a wide spread in $f_{quench}$ for different $B/T$ at a given $M_{Halo}$.  Naively, a different, less scattered behavior would be expected if halo mass quenching were the sole and dominant mechanism (Birnboim et al. 2007; Dekel \& Birnboim 2008; Dekel et al. 2009).  As a function of $M_{BH}$, on the other hand, a similar upturn in $f_{quench}$ is present for all $B/T$ bins equally above $\log(M_{BH}) = 7.5$.

The increased scatter in the relation between $f_{quench}$ and $M_{Bulge}$ compared to the tight correlation of $f_{quench}$ and  $M_{BH}$ can be explained by the SAM's $M_{BH}$ - $M_{Bulge}$ relation.  For a given $B/T$ bin, the scatter in the $M_{BH}$ - $M_{Bulge}$ relation is significant compared to the dynamic range in the bulge masses plotted.  At a given bulge mass, the scatter stems in part from an anti-correlation between the black hole mass and the level of star formation activity (sSFR).

Our analysis illustrates that in the Somerville model, which includes feedback from both supernovae and AGN, the central supermassive black hole acts as the primary agent of quenching in massive galaxies, and its accumulated mass (i.e., the integral over past accretion activity) is tightly related to the probability of finding a galaxy in a quenched state.  Since the physical processes giving rise to bulge and black hole growth are the same in the SAM (mergers and disk instabilities), the stronger relation of $f_{quench}$ with $M_{Bulge}$ than with $M_*$, present in both observations and model predictions, is not surprising.  

We stress that in the Somerville model, no direct causal link between the presence of a bulge and quenching is implemented.  The bulge is simply the accessible observable that correlates most tightly with the actual agent of quenching in this particular model: the supermassive black hole.  Observationally, there is increasing evidence for AGN-driven outflows of massive $z\sim 1-3$ galaxies (F\"{o}rster Schreiber et al.  2013; Nesvadba et al. 2011; Harrison et al. 2012; Cano-D{\'{\i}}az et al. 2012).  
Given the shortcomings of the SAM in a quantitative sense, notably its underprediction of the quiescent population at $z \sim 2$, additional or other quenching processes may be at play in the real universe.  One such process, that is causally linked to the presence of a bulge, could be morphological quenching (Martig et al. 2009; Genzel et al. 2014).  In such a scenario, the high central stellar density provided by a bulge, stabilizes the gas disk and prevents it from forming stars.  While plausibly only a temporary measure (as no gas is expelled, nor stopped from accreting through this mechanism), it could potentially contribute to suppressing star formation in $z \sim 2$ galaxies more efficiently and/or preventing them from returning to the star-forming branch in the SAM.  In the local universe, star formation efficiencies of bulge-dominated systems are reduced by factors of $\sim 2 -3$ compared to disk-dominated galaxies (Saintonge et al. 2012; Martig et al. 2013).

\section {Discussion}
\label{discussion.sec}

\subsection {Structural change}

In order to study the morphological differences between SFGs and quiescent galaxies, and to draw conclusions on the possible structural changes of star-forming galaxies as they move along the MS, we first examined the mass dependence of the profile shape of SFGs and quiescent galaxies as traced by the \sersic\ index and $B/T$ ratio.  We have shown that quiescent galaxies are structurally distinct from the star-forming population as seen by overall higher \sersic\ indices and $B/T$ ratios at a given stellar mass. SFGs show rising trends of their median \sersic\ index and $B/T$ ratio with increasing stellar mass, with the latter rising up to $\sim 40-50 \%$ above $10^{11}\ M_{\sun}$.  These findings give insights about the link between the structural evolution of SFGs and the quenching process as they move along the MS.

Analyzing the Schechter functional forms of the SFG and QG mass function as a function of redshift, \cite{peng2010} conclude that the quenching rate of galaxies climbing the MS rises proportionally to the SFR (and given the near-linear MS slope therefore also proportionally to the stellar mass, hence their terminology 'mass quenching').  This corresponds to a survival probability on the MS that drops exponentially with mass, implying that, while nearly all low-mass SFGs are destined to continue growing along the MS, toward the high-mass end the MS becomes progressively more dominated by near-to-be-dead SFGs.  In fact, the sub-unity slope of the MS, and possible flattening at the high-mass end (Whitaker et al. 2012), may well be interpreted in this context: the typical SFG above $10^{11}\ M_{\sun}$ is already undergoing some level of quenching, thereby deviating from the projected path along a SFR-Mass relation of slope unity, that could be expected from cosmological accretion rates in the absence of 
quenching.  Tying in our observational results on galaxy structure, the deviation toward high median $n$ and $B/T$ at the massive end reflects the typical structure of soon-to-be-dead star-formers that account for the bulk of SFGs above $10^{11}\ M_{\sun}$.  The fact that they are structurally distinct implies that the morphological transition happens first, to be followed later by the departure from the MS.  Bulge growth precedes quiescence.  
Such a morphological change prior to quenching is in line with qualitative predictions based on a toy model by Dekel \& Burkert (2013).  In the latter study, about half of the star-forming disk galaxies at $z\sim2$ are predicted to evolve into compact star-forming 'blue nuggets' due to violent disk instabilities before they are quenched into compact quiescent galaxies ('red nuggets').  An observed population of 'blue nuggets' has been proposed by Barro et al. (2013a,b) to represent an evolutionary link, originating from extended disk galaxies, and evolving into compact quiescent systems.

This does not refute that galaxies also undergo further structural evolution after they are quenched.  At least part of the size growth (Cassata et al. 2013; van der Wel et al. 2014) and evolution toward rounder axial ratios (Chang et al. 2013) has been attributed to (minor and/or major) dry mergers, and it is conceivable that similar processes contribute to the observed trend of increasing $B/T$ toward the massive end for the quiescent population.

\subsection {AGN as the driver of quenching ?}

We have demonstrated that the bulge mass of a system is well correlated with its quenched state and has a stronger predictive power of quiescence than the total stellar mass.  The observed trends of $f_{quench}$ with total stellar mass, bulge mass and disk mass as viewed among galaxies in different $B/T$ bins are in good qualitative agreement with predictions from the Somerville et al. SAM.  In the context of this model, the growth of the central supermassive black hole, which is the primary quenching agent for massive galaxies in this SAM, is tightly coupled with the growth of bulges through both merging and disk instabilities.

If a black hole - bulge scaling relation is in place during the peak of cosmic star formation as it is in the present-day universe (H\"aring \& Rix 2004), our observational results together with the model comparison could therefore hint at the bulge not being the causal link to quenching, but rather the most accessible observational proxy for the AGN acting as the quenching agent\footnote{We note that Rosario et al. (2013) find X-ray signatures of AGN activity at these high redshifts to be most prominent among the star-forming population, most notably at the high-mass end, precisely where we see an upturn in the bulge fraction among SFGs.}.  In detail, however, there are quantitative differences between the SAM and our observations, most severely in the highest redshift bin ($1.5 < z < 2.5$), where the observed quenched fraction exceeds the value predicted by the SAM by a factor of $\sim 3.5$.  The latter difference could hint at a need for more frequent, efficient, or lasting quenching, a possible mechanism 
we speculate about below.  We also note that the same behavior is not necessarily a generic feature to all SAMs (see Appendix C).

It is tempting to draw connections between the emerging bulges in massive MS galaxies out to $z \sim 2.5$ revealed by our analysis, and recent observational results based on deep AO-assisted integral field data sets and grism spectroscopy over the same redshift range.  F\"{o}rster Schreiber et al. (2013) found a high prevalence of powerful nuclear outflows in $\log(M) > 11$ galaxies driven by AGN, which appear to be absent in galaxies at lower masses.  Along with star formation driven winds in the outer parts of the galaxies, such outflows could efficiently remove gas out of galaxies and, in this way, contribute to the quenching process.

Meanwhile, the 3D-HST and CANDELS legacy programs have yielded evidence for nuclear depressions in the H$\alpha$ equivalent width in $z \sim 1$ SFGs (Nelson et al. 2012, 2013; Wuyts et al. 2013).  At the highest stellar surface mass densities, star formation no longer appears to proceed in lockstep with the assembled stellar mass.  Likewise, Genzel et al. (2014) report on ring-shaped H$\alpha$ distributions in $z \sim 2$ SFGs, surrounding a more quiescent center where the dynamically inferred Toomre Q parameter significantly exceeds unity, owing to the emergence of a stellar bulge.  As such, the Toomre stability criterion is satisfied in the central galaxy regions, which consequently could prevent the gas reservoir, if present there, from fragmenting and forming stars.  While this result suggests that some causal connection between bulge growth and quenching may be at play, it should be noted (as is done also by Genzel et al. 2014) that gravitational quenching by itself does not expel the gas present, 
neither does it stop the accumulation of a larger gas reservoir by continuing cosmological accretion.  Additional maintenance mode might be required for a long-term shut-down of further gas supply.

\section {Conclusions}
\label{conclusions.sec}

We analyzed the structural properties of a sample of 6764 massive ($ > 10^{10}\ M_{\sun}$) galaxies in the redshift range $ 0.5 < z < 2.5$, by exploiting the multi-wavelength CANDELS HST imaging data set in all five CANDELS/3D-HST fields.  We carried out single-component (\sersic) fits and two-component (bulge + disk) decompositions, on stellar mass maps reconstructed from a resolved panchromatic SED modeling technique (Wuyts et al. 2012, 2013), as well as on images of the $H$-band surface brightness distribution. 
In addition, we compared our findings to predictions by the state-of-the-art semi-analytic model from Somerville et al. (2008, 2012; with extensions including disk instabilities presented by Porter et al. 2014).  Our main results are the following:

\begin{enumerate}[label=\arabic*.,align=left, leftmargin=*]

\item At fixed stellar mass, quiescent galaxies have overall higher \sersic\ indices and $B/T$ ratios than SFGs as measured on their mass maps, in line with previous findings using monochromatic observations.  We find an increase of \sersic\ indices among SFGs with increasing total stellar mass, with the median mass profiles increasing from ($n\sim1.3$) at $10^{10}\ M_{\sun}$, to $n \gtrsim 2$ above $10^{11}\ M_{\sun}$.  Two-component bulge-disk decompositions confirm that the same rising trend is present when considering the median $B/T$ ratio of SFGs, which is rising up to $\sim 40-50 \%$ above $10^{11}\ M_{\sun}$.  The same characteristic behavior is seen at $z \sim 1$ and $z \sim 2$. 

\item Quantifying the same trends on the H-band light profiles rather than the mass maps, the \sersic\ indices and $B/T$ fractions are overall lower for SFGs, confirming previous non-parametric measurements for a subset of our sample (Wuyts et al. 2012).  The emergence of bulges above $10^{11}\ M_{\sun}$ in SFGs appears to be also slightly less prominent when viewed in light, consistent with the steepest color gradients (blue disks with red central bulges) being found among massive SFGs.

\item The likelihood of a galaxy being quenched, as traced by the fraction of quiescent galaxies, is better correlated with the bulge mass than the total stellar mass and further shows no appreciable correlation with the amount of stellar mass in the disk component.  The quenched fraction at redshift 1 is on average higher by a factor $\sim 2$ than at redshift 2. 

\item At a given total stellar mass, the quenched fraction exhibits a strong positive correlation with $B/T$, while different $B/T$ bins are confined to a significantly tighter locus in a diagram of $f_{quench}$ versus $M_{Bulge}$.  These findings imply that the bulge mass of a system is the single observable parameter with the most predictive power regarding its quenched state, although a somewhat tighter constraint on the probability of quiescence can be obtained if in addition also the total stellar mass is known.  The same trend is seen over the full redshift range probed, with the distinction that quenched fractions are lower at higher lookback times.  

\item We find a good qualitative agreement between the semi-analytic model by Somerville et al. SAM and our observational findings.  Since bulge and black hole growth are tightly coupled in the SAM, the strong dependence of $f_{quench}$ on bulge mass follows rather naturally in this model.  Our observational results can {\it in the context of this model} therefore be interpreted as the bulge being the closest observable proxy to the underlying agent of quenching: the black hole.  Quantitatively, the largest discrepancy between model and observations is found in the highest redshift bin ($1.5 < z < 2.5$), where the observed quenched fraction is larger by a factor of $\sim 3.5$ than predicted by the SAM.  We note that the same behavior is not necessarily a generic feature to all SAMs.

\end{enumerate}

\section*{}

The authors acknowledge fruitful discussions with Edmond Cheung, David C. Koo, Yu Lu, Casey J.  Papovich, Mohammadtaher Safarzadeh, Benjamin J. Weiner and Steven P. Willner. Support
for Program number HST-GO-12060 and HST-GO-12177 was
provided by NASA through a grant from the Space Telescope
Science Institute, which is operated by the Association of
Universities for Research in Astronomy, Incorporated, under
NASA contract NAS5-26555.

\vspace{10mm}
\begin {appendix}
\section {A. The meaning of a Sersic index measurement}
\label{A.app}

\begin {figure*}[htp]
\centering
 \includegraphics[width=\textwidth]{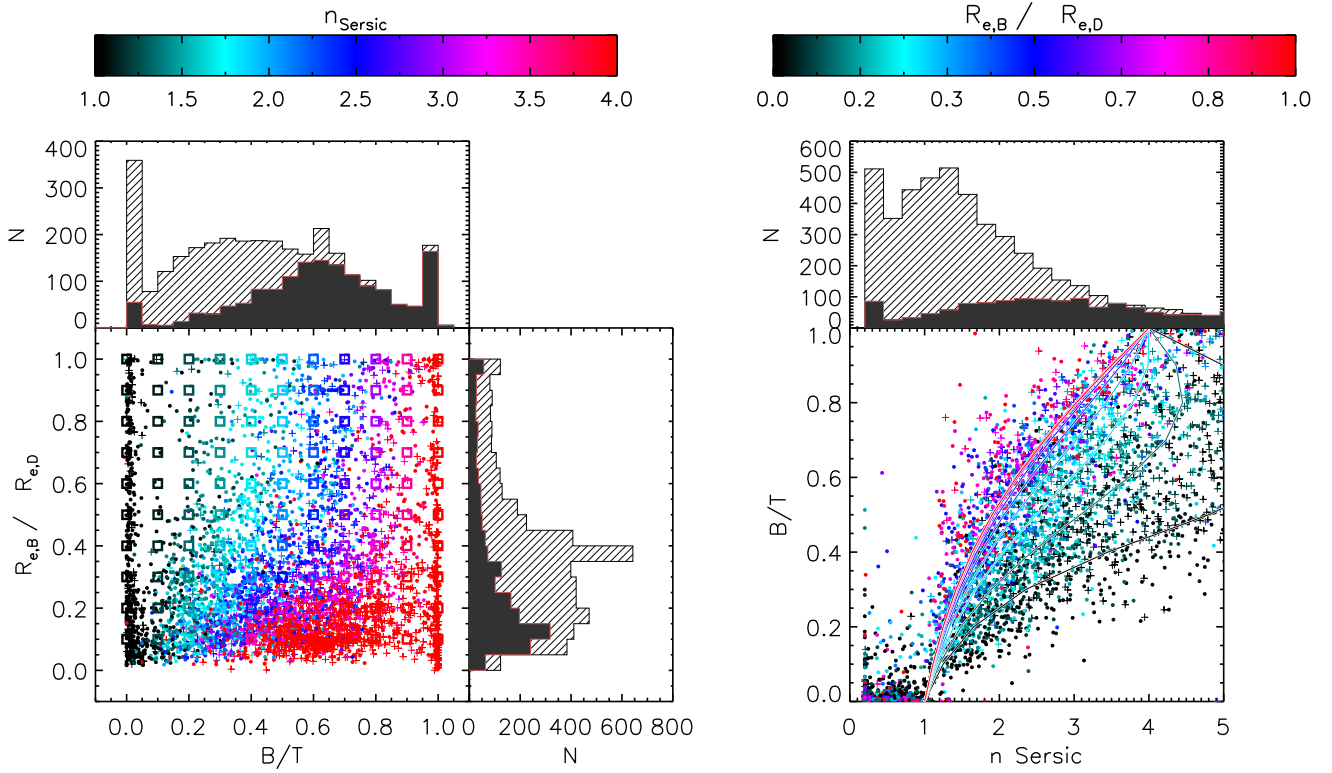}
\caption{Left: Size ratio between bulge and disk component ($R_{e,B}/R_{e,D}$) vs. $B/T$ for the whole galaxy sample as measured in mass.  The color indicates the measured \sersic\ index. Quiescent galaxies are shown as a plus (+), SFGs are shown as dots.
Right:  Measured $B/T$ values against \sersic\ indices as seen in mass.  The histograms show the respective measured distributions of $B/T$, $R_{e,B}/R_{e,D}$ and $n$ among the sample for quiescent galaxies (solid black areas) and SFGs (hatched areas).  
The relationship between the \sersic\ index and $B/T$ is not unique, but rather crucially depends on $R_{e,B}/R_{e,D}$.  The squares (left) and curved lines (right) illustrate the relation between ($B/T,R_{e,B}/R_{e,D}$) and $n$ as inferred from noise free idealized bulge + disk toy models, consistent with the trends seen for our observed galaxy samples.
\label{n_bd.fig}}
\end {figure*}

The combination of our one and two-component fits on the mass maps enables us to examine empirically how the \sersic\ index relates the amplitude of the bulge (as parametrized by $B/T$) on the one hand, and the relative size-ratio to the bulge and disk component ($R_{e,B}/R_{e,D}$) on the other hand.  Figure\ \ref{n_bd.fig} illustrates that, while $B/T$ shows a clear correlation with $n$, there is no unique one-to-one translation between the two.  Instead, the best-fit $n$ to a composite bulge+disk system additionally depends on the size ratio of the two components.  In other words, a galaxy's Sersic index could be increased by placing more material in the bulge, but also by growing the disk at fixed bulge size.  The observed CANDELS galaxies occupy a surface in this three-parameter space ($n$, $B/T$, $R_{e,B}/R_{e,D}$) that is in good agreement with what would be anticipated from Sersic fits to idealized, noise-free bulge ($n = 4$) plus disk ($n = 1$) profiles (squares and curves 
in the left-and right-hand panels of Figure\ \ref{n_bd.fig}, respectively).  
Evidently, for systems with $B/T$ close to 0 or 1, the size ratio of the two components is ill-constrained as one of them contains barely any mass.  The galaxies in our full $0.5 < z < 2.5$ sample span the full range of $B/T$ values, and are located predominantly around bulge-to-disk size ratios of $R_{e,B}/R_{e,D} \sim 0.2$ .

\section {B. Comparison with Measurements on H-band}
\label{B.app}

\begin {figure*}[htp]
\centering
\vspace{-6mm}
 \includegraphics[width=0.49\textwidth]{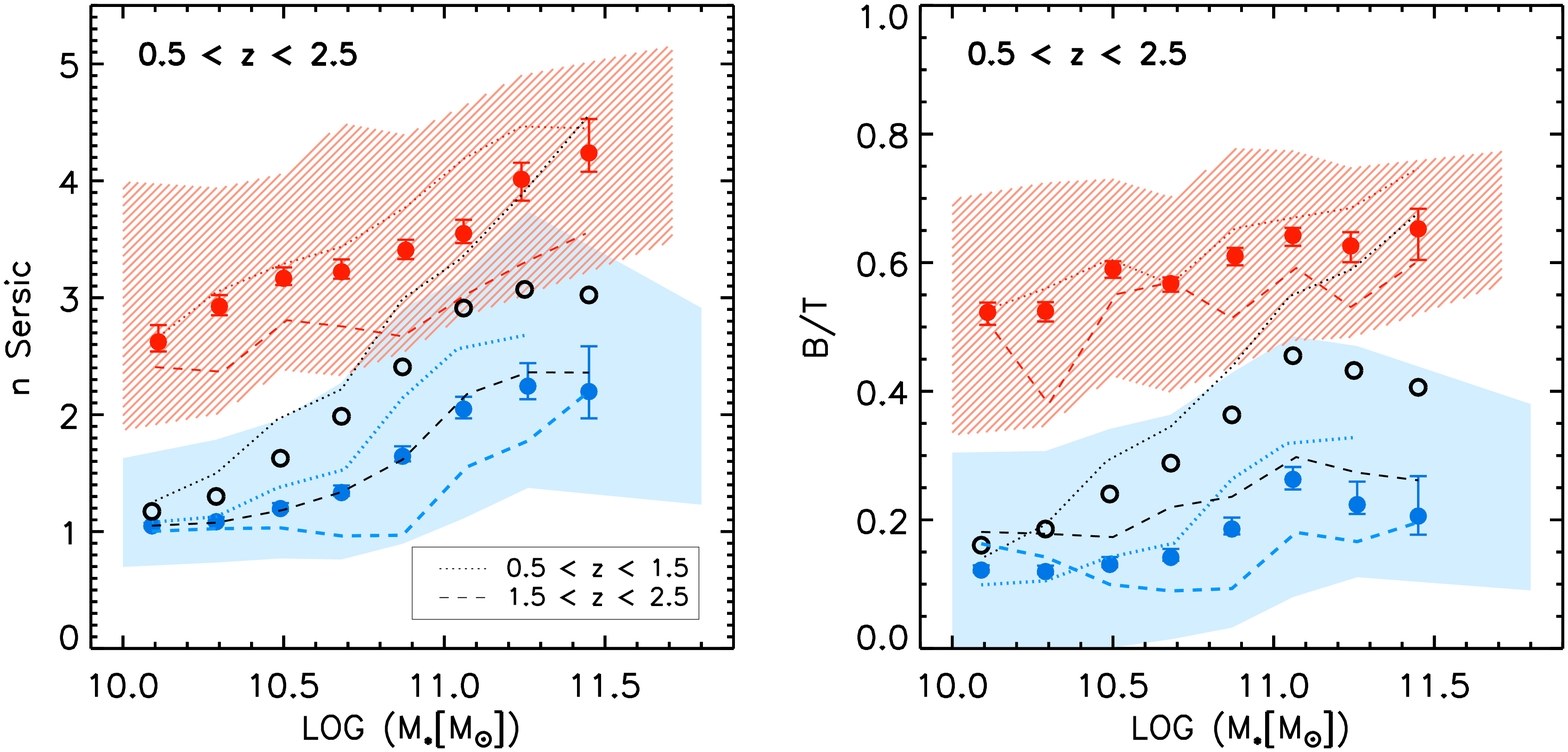}
 \includegraphics[width=0.49\textwidth]{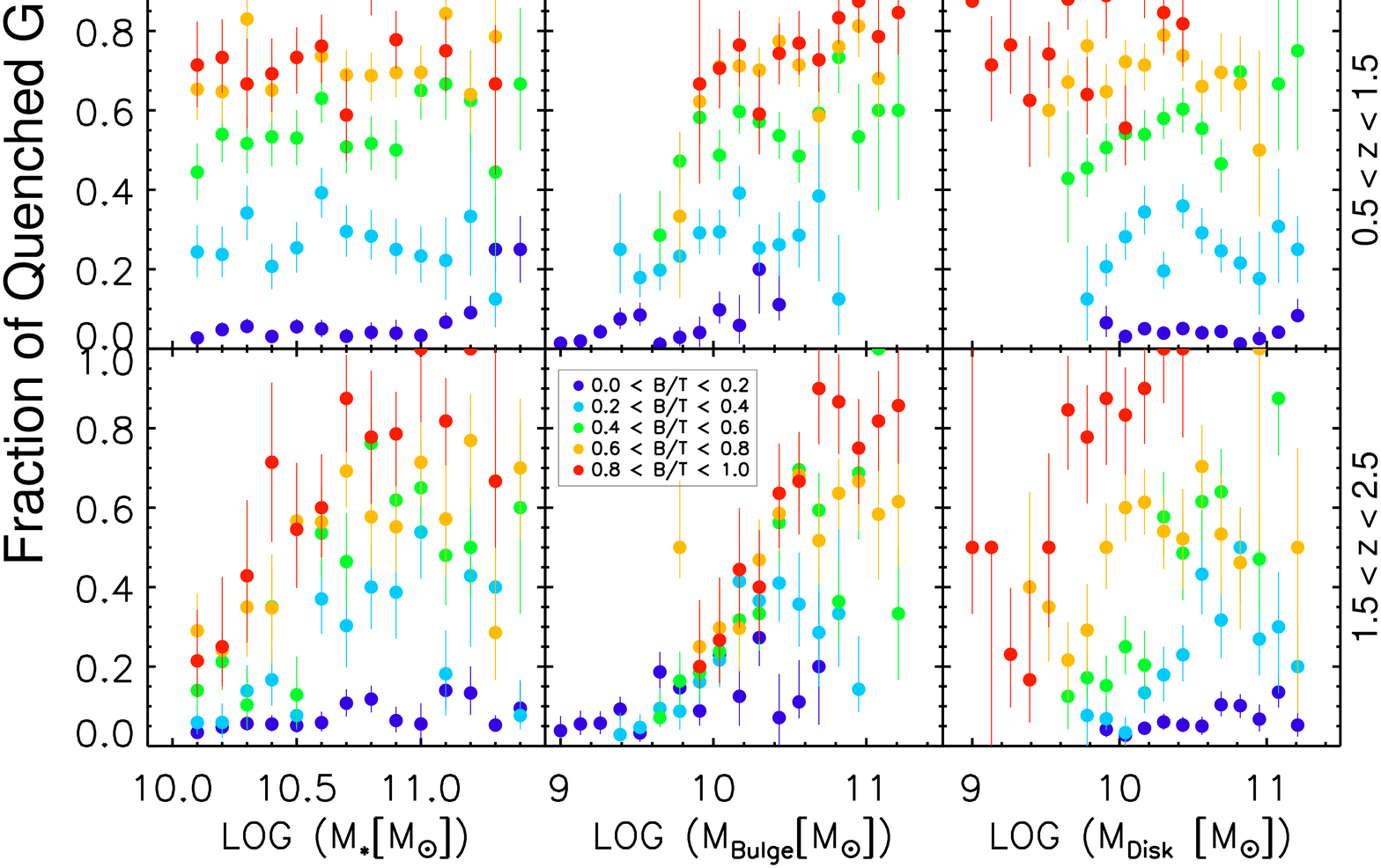}
\caption{Equivalent of Figures\ \ref{struc.fig} (Left) and\ \ref{fquench_all.fig} (right) using the results of structural measurements on the $H$-band light images.
\label{light_struc.fig}}
\end {figure*}

Here, we investigate how the results of our structural analysis change when conducting the measurements on the $H$-band light images rather than on the stellar mass maps.  To this end, we present in Figure\ \ref{light_struc.fig} the same figures as discussed in Section\ \ref{structure.sec}, now using the \sersic\ index and $B/T$ values as inferred from the CANDELS $H$-band imaging.  Likewise, we compute the bulge and disk mass as $(B/T)_H M_*$ and $(1 - (B/T)_H) M_*$, respectively.

Overall, our analysis reveals a qualitatively similar mass dependence of $n$ and $B/T$, and distinction between SFGs and quiescent galaxies as inferred from the mass maps.  In detail, however, modest changes in $n$ and $B/T$ are notable.  While the median $z \sim 1$ (2) SFG has $(B/T)_H \lesssim 0.25$ (0.20), the typical bulge fractions increase to above 20\%, and reach up to $\sim 40 - 50\%$ at the massive end, once spatial $M/L$ variations are corrected for.  Likewise, the corresponding \sersic\ indices measured  on $H$-band imaging for SFGs below $\log(M) = 10.8$ are consistent with exponential disk profiles (see also Wuyts et al. 2011), but are slightly cuspier as quantified on mass maps.  This is in line with findings based on smaller subsets of CANDELS data by Wuyts et al. (2012) and Guo et al. (2012).

The central 50th percentile intervals marked by the red and blue polygons are somewhat less confined in the plots based on stellar mass maps compared to the $H$-band results.  We interpret this to be due to an additional source of random uncertainty introduced by the resolved stellar population modeling.  The resolved stellar population modeling itself was motivated by the need to reduce the systematic biases associated with spatial $M/L$ ratio variations. 

Focussing on Figure\ \ref{light_struc.fig}, the qualitative trends of $f_{quench}$ with total stellar mass, bulge and disk mass are very similar.  However, the lower two rows of Figure\ \ref{light_struc.fig} compared to Figure\ \ref{fquench_all.fig} show that the bins of lowest and highest $B/T$ are more separated from each other in $f_{quench}$ in light than in mass.  This observation too is in line with the disks of SFGs being dominating by a younger stellar population than that of the bulge, shifting SFGs to lower $B/T$.  The middle panels of Figure\ \ref{light_struc.fig} show a larger scatter than the corresponding panels in Figure\ \ref{fquench_all.fig}. 

\section {C. Comparison to Guo et al. (2013) SAM}
\label{C.app}

\begin {figure*}[htp]
\centering
  \includegraphics[width=0.585\textwidth]{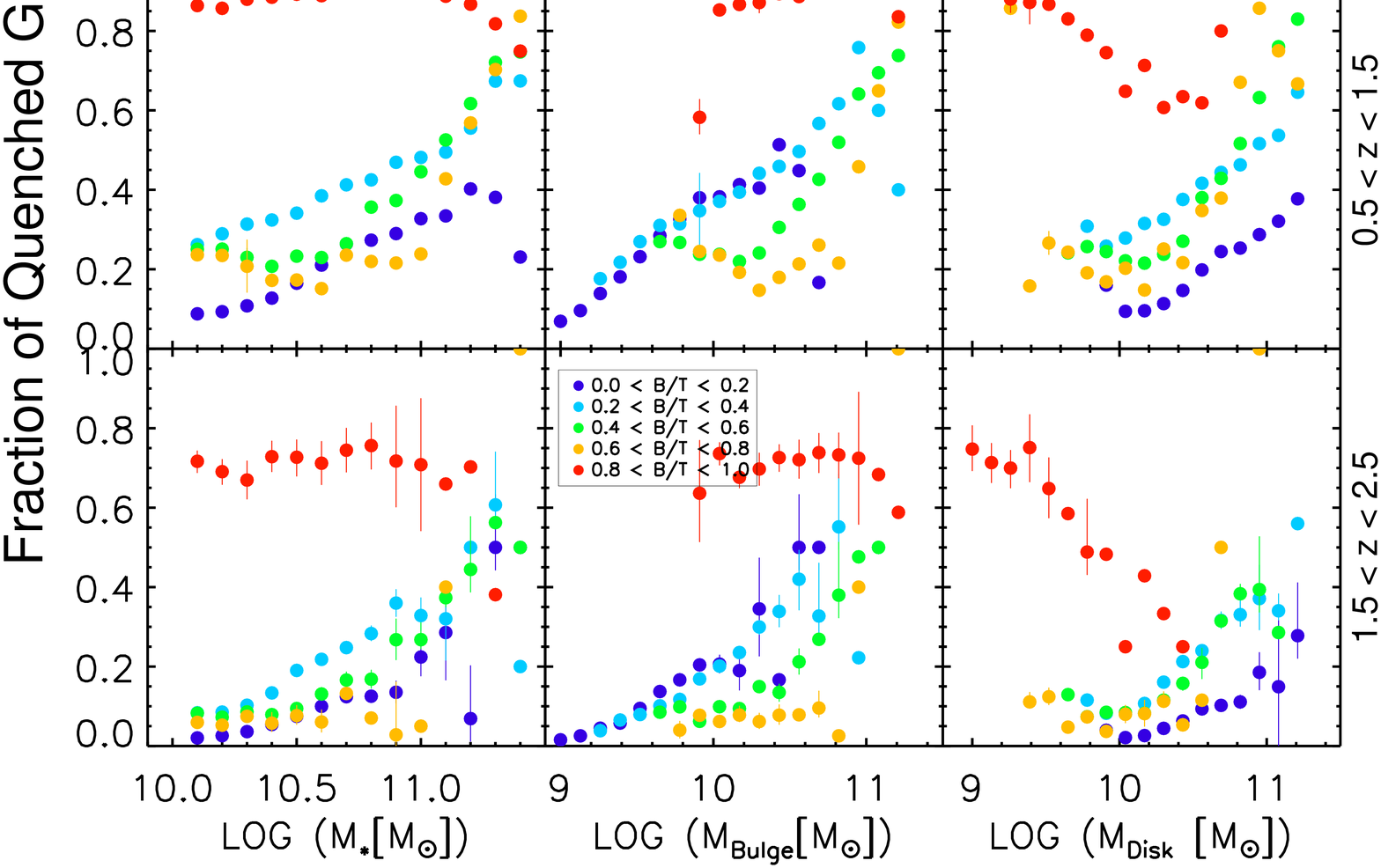}
 \includegraphics[width=0.41\textwidth]{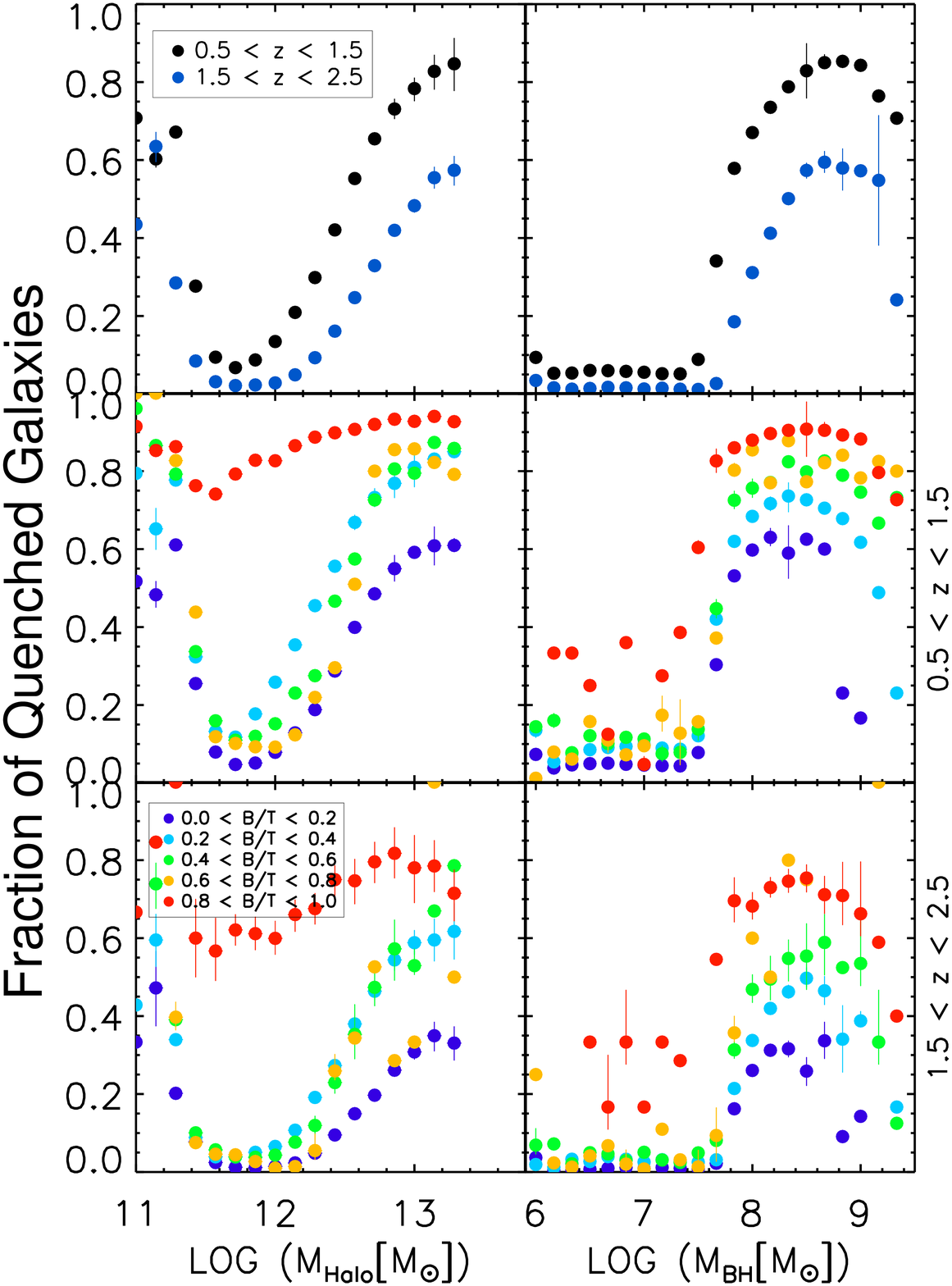}
\caption{Fraction of massive galaxies ($M > 10^{10}\ M_{\sun} $) that are quiescent in the SAM of Guo et al. (2013), as a function of total stellar mass, stellar mass in the bulge and disk separately, and as a function of halo and black hole mass.  The bottom two rows show the results split by bulge-to-total ratio, for the $z \sim 1$ and $z \sim 2$ bins respectively.
\label{SAM_Guo.fig}}
\end {figure*}

In order to evaluate to which extent the characteristic trends of the 
Somerville et al. SAM, as presented in Section\ \ref{SAM_results.sec}, are 
generic to all semi-analytic models, we considered an independent 
semi-analytic model by Guo et al. (2013).  The Guo et al. (2013) model 
is rooted in the dark matter backbone of the Millennium Simulation 
(Springel et al. 2005), and its output tables with galaxy properties for 
snapshots of different lookback times are publicly available on the 
Virgo Millennium Database (G. Lemson \& the Virgo Consortium 2006).  We 
present the equivalent plots of Figure\ \ref{SAM_Somerville1.fig} and\ \ref{SAM_Somerville2.fig} in Figure\ 
\ref{SAM_Guo.fig}.  As in the Somerville model, quenched fractions are rising with 
increasing mass, increasing bulge mass, and decreasing redshift, but 
broken down in bins of $B/T$, significant differences are notable. 
Specifically, $f_{quench}$ does not monotonically increase with 
increasing $B/T$ at a given stellar mass.

\end {appendix}

\end {document}